\documentstyle[12pt,ssi]{article}

\def \MSbar {\vbox{\hrule\kern 1pt\hbox{\rm MS}}}
\def \GeV { {\ \rm GeV} }
\def \MeV { {\ \rm MeV} }
\def\lesssim{\ \hbox{\raise 2pt \hbox{$<$} \kern -13pt
                     \lower 3pt \hbox{$\sim$}}\ }
\def\red#1{#1}
\def\orange#1{#1}
\def\blue#1{#1}
\def\green#1{#1}
\def\sienna#1{#1}
\def\magenta#1{#1}

\input epsf.sty
\def\DESepsf(#1 width #2){\epsfxsize=#2 \epsfbox{#1}}

\begin{document}

\title{Basics of QCD Perturbation Theory}

\author{Davison E.\ Soper
\thanks{Supported by DOE Contract DE-FG03-96ER40969.}\\ 
Institute of Theoretical Science \\
University of Oregon, Eugene, OR 97403}

\maketitle

\begin{abstract}

\noindent
This is an introduction to the use of QCD perturbation theory,
emphasizing generic features of the theory that enable one to
separate short-time and long-time effects. I also cover some
important classes of applications: electron-positron annihilation to
hadrons, deeply inelastic scattering, and hard processes
in hadron-hadron collisions.

\vskip 10 cm

\hskip -1.5cm
{\it Lectures at the SLAC Summer Institute, August 1996}
\end{abstract}

\section{Introduction}

A prediction for experiment based on perturbative QCD combines
a particular calculation of Feynman diagrams with the use
of general features of the theory. The {\blue{particular
calculation}} is easy at leading order, not so easy at
next-to-leading order and extremely difficult beyond the
next-to-leading order. This calculation of Feynman
diagrams would be a purely academic exercise if we did not use
certain {\blue{general features}} of the theory that
allow the Feynman diagrams to be related to experiment:
\begin{itemize}
\item the renormalization group and the running
coupling;
\item the existence of infrared safe observables;
\item the factorization property that allows
us to isolate hadron structure in parton distribution
functions.
\end{itemize}

In these lectures, I discuss these structural features of the theory
that allow a comparison of theory and experiment. Along the way we
will discover something about certain important processes:
\begin{itemize}
\item ${e^+e^-}$ annihilation;
\item deeply inelastic scattering;
\item hard processes in hadron-hadron collisions.
\end{itemize}
By discussing the particular along with the general, I hope to arm
the reader with information that speakers at research conferences take
to be collective knowledge -- knowledge that they assume the audience
already knows.

Now here is the {\sienna{disclaimer}}. We will not learn how
to do significant calculations in QCD perturbation theory. Three
lectures is not enough for that.

I hope that the reader may be inspired to pursue the subjects
discussed here in more detail. A good source is the {\it Handbook of
Perturbative QCD}\cite{handbook} by the CTEQ collaboration. More
recently, Ellis, Stirling and Webber have written an excellent book
\cite{ESW} that covers the most of the subjects sketched in these
lectures. For the reader wishing to gain a mastery of the theory, I
can recommend the recent books on quantum field theory by Brown
\cite{BrownQFT}, Sterman \cite{StermanQFT},  Peskin and Schroeder
\cite{PeskinQFT}, and Weinberg \cite{WeinbergQFT}. Another good
source, including both theory and phenomenology, is the lectures in
the 1995 TASI proceedings, {\it QCD and Beyond}\cite{TASI}.

\section{Electron-positron annihilation and jets}

In this section, I explore the structure of the final state in QCD.
I begin with the kinematics of $e^+e^- \to 3\ partons$, then examine
the behavior of the cross section for $e^+e^- \to 3\ partons$ when
two of the parton momenta become collinear or one parton momentum
becomes soft. In order to illustrate better what is going on, I
introduce a theoretical tool, null-plane coordinates. Using this
tool, I sketch a space-time picture of the singularities that we
find in momentum space. The singularities of perturbation
theory correspond to long-time physics. We see that the structure of
the final state suggested by this picture conforms well with what is
actually observed.

I draw a the distinction between short-time physics, for which
perturbation theory is useful, and long-time physics, for which the
perturbative expansion is out of control.  Finally, I discuss how
certain experimental measurements can probe the short-time physics
while avoiding sensitivity to the long-time physics.

\subsection{Kinematics of $ e^+ e^- \to 3$ partons}

\begin{figure}[htb]
\centerline{\DESepsf(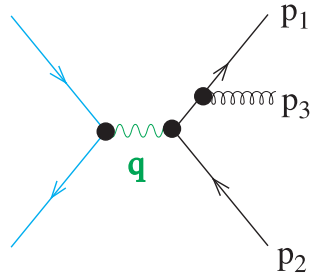 width 4 cm)}
\caption{Feynman diagram for $e^+e^- \to q\,\bar q\,g$.}
\label{eetojetsA}
\end{figure}

Consider the process $e^+e^- \to q\,\bar q\,g$, as illustrated in
Fig.~\ref{eetojetsA}. Let $\sqrt s$ be the total energy in the
c.m.\ frame and let $q^\mu$ be the virtual photon (or Z boson)
momentum, so $q^\mu q_\mu = s$. Let $p_i^\mu$ be the momenta of the
outgoing partons $(q,\bar q,g)$ and let $E_i = p_i^0$ be the
energies of the outgoing partons. It is useful to define energy
fractions $x_i$ by 
\begin{equation}
{\blue{x_i =  {E_i \over \sqrt s/2}}} = {2 p_i\cdot q\over s}.
\end{equation}
Then 
\begin{equation}
{\red{ 0<x_i}}.
\end{equation}
Energy conservation gives
\begin{equation}
{\red{\sum_i x_i}} =   {2(\sum p_i)\cdot q\over s} {\red{= 2}}.
\end{equation}
Thus only two of the $x_i$ are independent.

Let $\theta_{ij}$ be the angle between the momenta of partons $i$ and
$j$. We can relate these angles to the momentum fractions as follows:
\begin{equation}
2 p_1\cdot p_2 =
 (p_1 + p_2)^2=
 (q - p_3)^2 =
s - 2 q\cdot p_3,
\end{equation}
\begin{equation}
2 E_1 E_2 (1 - \cos \theta_{12}) = 
s (1-x_3).
\end{equation}
Dividing this equation by $s/2$ and repeating the argument for the
two other pairs of partons, we obtain three relations for the angles
$\theta_{ij}$:
\begin{eqnarray}
{\blue{x_1 x_2 (1 - \cos \theta_{12})}} &=& 
{\blue{2 (1-x_3)}},
\nonumber\\
{\blue{x_2 x_3 (1 - \cos \theta_{23})}} &=& 
{\blue{2 (1-x_1)}},
\nonumber\\
{\blue{x_3 x_1 (1 - \cos \theta_{31})}} &=& 
{\blue{2 (1-x_2)}}.
\end{eqnarray}
We learn two things immediately. First,
\begin{equation}
{\red{x_i <1}} .
\end{equation}
Second, the three possible collinear configurations of the partons
are mapped into $x_i$ space very simply:
\begin{eqnarray}
{\red{\theta_{12}\to 0}} &\Leftrightarrow& {\red{x_3 \to 1}},
\nonumber\\
{\red{\theta_{23}\to 0}} &\Leftrightarrow& {\red{x_1 \to 1}},
\nonumber\\
{\red{\theta_{31}\to 0}} &\Leftrightarrow& {\red{x_2 \to 1}}.
\end{eqnarray}

\begin{figure}[htb]
\centerline{\DESepsf(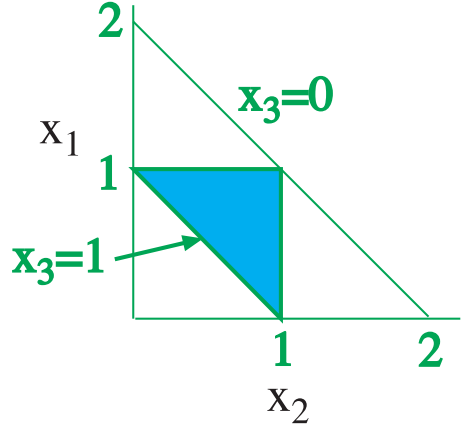 width 6 cm)}
\caption{Allowed region for $(x_1,x_2)$. Then $x_3$ is $2 - x_1 -
x_2$.}
\label{eetojetsB}
\end{figure}
 
\begin{figure}[htb]
\centerline{\DESepsf(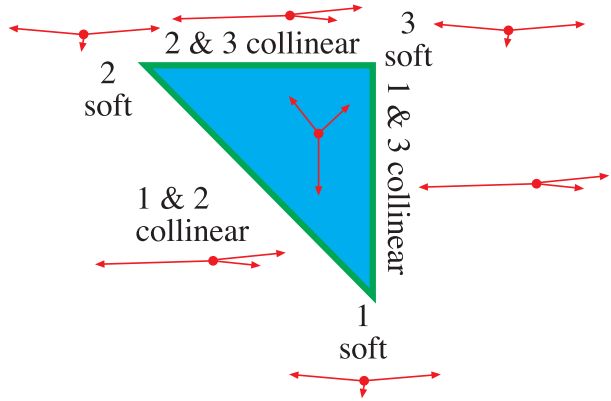 width 8 cm)}
\caption{Allowed region for $(x_1,x_2)$. The labels and small
pictures show the physical configuration of the three partons
corresponding to subregions in the allowed triangle.}
\label{eetojetsC}
\end{figure}

The relations $0 \le x_i \le 1$, together with $x_3 = 2 - x_1 -
x_2$, imply that the allowed region for $(x_1,x_2)$ is a triangle,
as shown in Fig.~\ref{eetojetsB}. The edges $x_i = 1$ of the allowed
region correspond to two partons being collinear, as shown in
Fig.~\ref{eetojetsC}. The corners $x_i = 0$ correspond to one parton
momentum being soft ($p_i^\mu \to  0$).

\subsection{Structure of the cross section}

One can easily calculate the cross section corresponding to 
Fig.~\ref{eetojetsA} and the similar amplitude in which the gluon
attaches to the antiquark line. The result is
\begin{equation}
{1 \over \sigma_0}\,{d \sigma \over d x_1 dx_2}
= {\alpha_s \over 2\pi}C_F
{x_1^2 + x_2^2 \over {\red{(1 - x_1)(1 - x_2)}}},
\end{equation}
where $C_F = 4/3$ and $\sigma_0 = (4 \pi \alpha^2/s)\sum Q_f^2$ is the
total cross section for $e^+ e^- \to hadrons$ at order $\alpha_s^0$.
The cross section has collinear singularities:
\begin{eqnarray}
(1 - x_1) &\to& 0\,, \hskip 1 cm {\rm (2\&3\ collinear)};
\nonumber\\
(1 - x_2) &\to& 0\,, \hskip 1 cm {\rm (1\&3\ collinear)}.
\end{eqnarray}
There is also a singularity when the gluon is soft: $x_3 \to 0$. In
terms of $x_1$ and $x_2$, this singularity occurs when
\begin{equation}
(1 - x_1) \to 0,\quad (1 - x_2) \to 0,\quad 
{(1 - x_1) \over (1 - x_2) } \sim const.
\end{equation}

Let us write the cross section in a way that displays the collinear
singularity at $\theta_{31} \to 0$ and the soft singularity at
$E_3\to 0$:
\begin{equation}
{1\over \sigma_0}{d \sigma \over dE_3\, d \cos\theta_{31}}
= {\alpha_s \over 2\pi}\, C_F\,
{{\green{f(E_3,\theta_{31})}}
\over
{\red{E_3(1 - \cos \theta_{31})}}}.
\end{equation}
Here ${\green{f(E_3,\theta_{31})}}$ a rather complicated function.
The only thing that we need to know about it is that it is finite for
$E_3 \to 0$ and for $\theta_{31}\to 0$.

Now look at the collinear singularity, ${\red{\theta_{31}\to 0}}$.
If we integrate over the singular region holding $E_3$ fixed we find
that the integral is divergent:
\begin{equation}
\int_{a}^1d\cos\theta_{31}\
{d \sigma \over dE_3\, d \cos\theta_{31}}
= \log(\infty).
\end{equation}
Similarly, if we integrate over the region of the soft singularity,
holding $\theta_{31}$ fixed, we find that the integral is divergent:
\begin{equation}
\int_0^a dE_3\
{d \sigma \over dE_3\, d \cos\theta_{31}}
= \log(\infty).
\end{equation}
Evidently, perturbation theory is telling us that we should not take 
the perturbative cross section too literally. The total cross
section for $e^+e^- \to hadrons$ is certainly finite, so this partial
cross section cannot be infinite. What we are seeing is a breakdown
of perturbation theory in the soft and collinear regions, and we
should understand why.

\begin{figure}[htb]
\centerline{\DESepsf(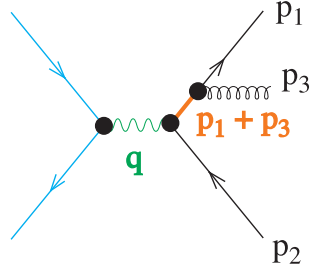 width 4 cm)}
\caption{Cross section for $e^+e^- \to q\,\bar q\,g$, illustrating the
singularity when the gluon is soft or collinear with the quark.}
\label{eetojetsD}
\end{figure}

Where do the singularities come from? Look at Fig.~\ref{eetojetsD}
(in a physical gauge). The scattering matrix element ${\cal M}$
contains a factor $1/{\orange{(p_1 + p_3)}}^2$ where
\begin{equation}
(p_1 + p_3)^2 = 2 p_1 \cdot p_3 = 2 E_1 E_3 (1 - \cos \theta_{31}).
\end{equation}
Evidently, $1/{\orange{(p_1 + p_3)}}^2$ is singular when
$\theta_{31} \to 0$ and when $E_3 \to 0$. The collinear singularity
is somewhat softened because the numerator of the Feynman diagram
contains a factor proportional to $\theta_{31}$ in the collinear
limit.  (This is not exactly obvious, but is easily seen by
calculating. If you like symmetry arguments, you can derive this
factor from quark helicity conservation and overall angular momentum
conservation.) We thus find that
\begin{equation}
|{\cal M}|^2 \propto \left[{\theta_{31} \over E_3
\theta_{31}^2}\right]^2
\end{equation}
for $E_3 \to 0$ and $\theta_{31} \to 0$. Note the universal nature of
these factors.

Integration over the double singular region of the momentum space for
the gluon has the form
\begin{equation}
\int {E_3^2d E_3 d \cos\theta_{31} d \phi \over E_3}
\sim 
\int E_3 d E_3 d \theta_{31}^2 d \phi .
\end{equation}
Combining the integration with the matrix element squared gives
\begin{equation}
d \sigma \sim 
\int E_3 d E_3 d \theta_{31}^2 d \phi
\left[{\theta_{31} \over E_3 \theta_{31}^2}\right]^2 \sim
\int  {{\red{d E_3}}\over {\red{E_3}}}\ 
{ {\red{d \theta_{31}^2}} \over {\red{\theta_{31}^2}}}\ d \phi.
\end{equation}
Thus we have a double logarithmic divergence in perturbation theory
for the soft and collinear region. With just a little enhancement of
the argument, we see that there is a collinear divergence from
integration over $\theta_{31}$ at finite $E_3$ and a separate soft
divergence from integration over  $E_3$ at finite $\theta_{31}$.
Essentially the same argument applies to more complicated graphs.
There are divergences when two final state partons become collinear
and when a final state gluon becomes soft. Generalizing further
\cite{Stermanpinch}, there are also divergences when several final
state partons become collinear to one another or when several (with
no net flavor quantum numbers) become soft.  

We have seen that if we integrate over the singular region in
momentum space with no cutoff, we get infinity. The integrals are
logarithmically divergent, so if we integrate with an infrared cutoff
$M_{IR}$, we will get big logarithms of $M_{IR}^2/s$.  Thus the
collinear and soft singularities represent perturbation theory out of
control. Carrying on to higher orders of perturbation theory, one gets
\begin{equation}
1 + \alpha_s  \times ({\red{{\rm big}}}) +
\alpha_s^2  \times ({\red{{\rm big}}})^2 + \cdots.
\label{outofcontrol}
\end{equation}
If this expansion is in powers of $\alpha_s(M_Z)$, we have $\alpha_s
\ll 1$. Nevertheless, the big logarithms seem to spoil any chance of
the low order terms of perturbation theory being a good approximation
to any cross section of interest. Is the situation hopeless? We
shall have to investigate further to see.

\subsection{Interlude: Null plane coordinates}
\label{NullPlane}

\begin{figure}[htb]
\centerline{\DESepsf(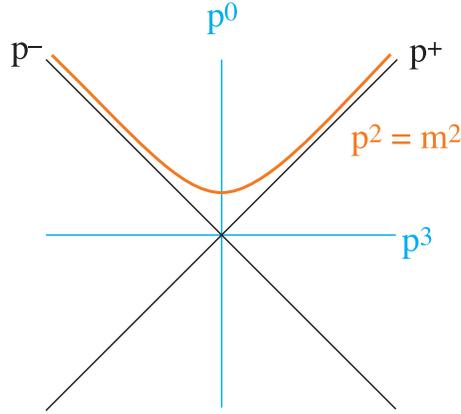 width 6 cm)}
\caption{Null plane axes in momentum space.}
\label{nullplane}
\end{figure}

In order to understand better the issue of singularities, it is
helpful to introduce a concept that is generally quite useful in high
energy quantum field theory, null plane coordinates. The idea is to
describe the momentum of a particle using momentum components $p^\mu =
(p^+,p^-,p^1,p^2)$ where
\begin{equation}
{\blue{p^\pm = (p^0 \pm p^3)/\sqrt 2}}.
\end{equation}
For a particle with large momentum in the $+z$ direction and limited
transverse momentum, $p^+$ is large and $p^-$ is small. Often one
{\it chooses} the plus axis so that a particle or group of particles
of interest have large $p^+$ and small $p^-$ and $p_T$.

Using null plane components, the covariant square of $p^\mu$ is
\begin{equation}
p^2 = 2 p^+p^- - { \bf p}_T^2.
\end{equation}
Thus, for a particle on its mass shell, $p^-$ is
\begin{equation}
p^- = {{ \bf p}_T^2 + m^2 \over 2p^+}.
\end{equation}
Note also that, for a particle on its mass shell,
\begin{equation}
p^+ > 0\,, \hskip 1 cm p^- >0\,.
\end{equation}
Integration over the mass shell is
\begin{equation}
(2\pi)^{-3} \int {d^3\vec p \over 2 \sqrt{\vec p^2 + m^2}}\cdots
= (2\pi)^{-3}
\int\! d^2{\bf p}_T\, \int_0^\infty\!{dp^+ \over 2 p^+}\cdots.
\end{equation}

We also use the plus/minus components to describe a space-time point
$x^\mu$: ${\blue{x^\pm = (x^0 \pm x^3)/\sqrt 2}}$. In describing a
system of particles moving with large momentum in the plus direction,
we are invited to think of $x^+$ as ``time.'' Classically, the
particles in our system follow paths nearly parallel to the $x^+$
axis, evolving slowly as it moves from one $x^+ = const.$ plane to
another.

We relate momentum space to position space for a quantum system by
Fourier transforming. In doing so, we have a factor $\exp(i p\cdot
x)$, which has the form
\begin{equation}
{\blue{p\cdot x = p^+ x^- + p^- x^+ - {\bf p}_T \cdot{\bf x}_T}}.
\end{equation}
Thus $x^-$ is conjugate to $p^+$ and $x^+$ is conjugate to $p^-$.
That is a little confusing, but it is simple enough.

\subsection{Space-time picture of the singularities}

\begin{figure}[htb]
\centerline{\DESepsf(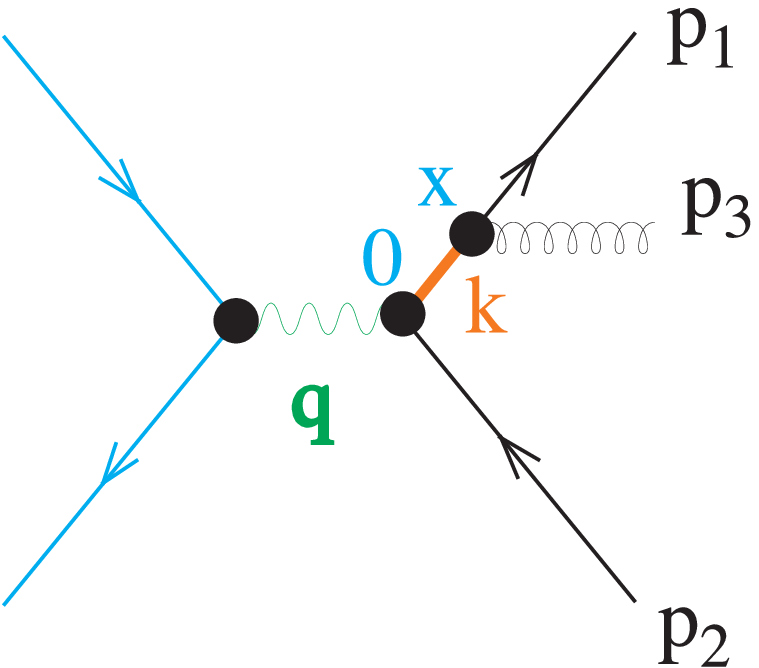 width 6 cm)
            \DESepsf(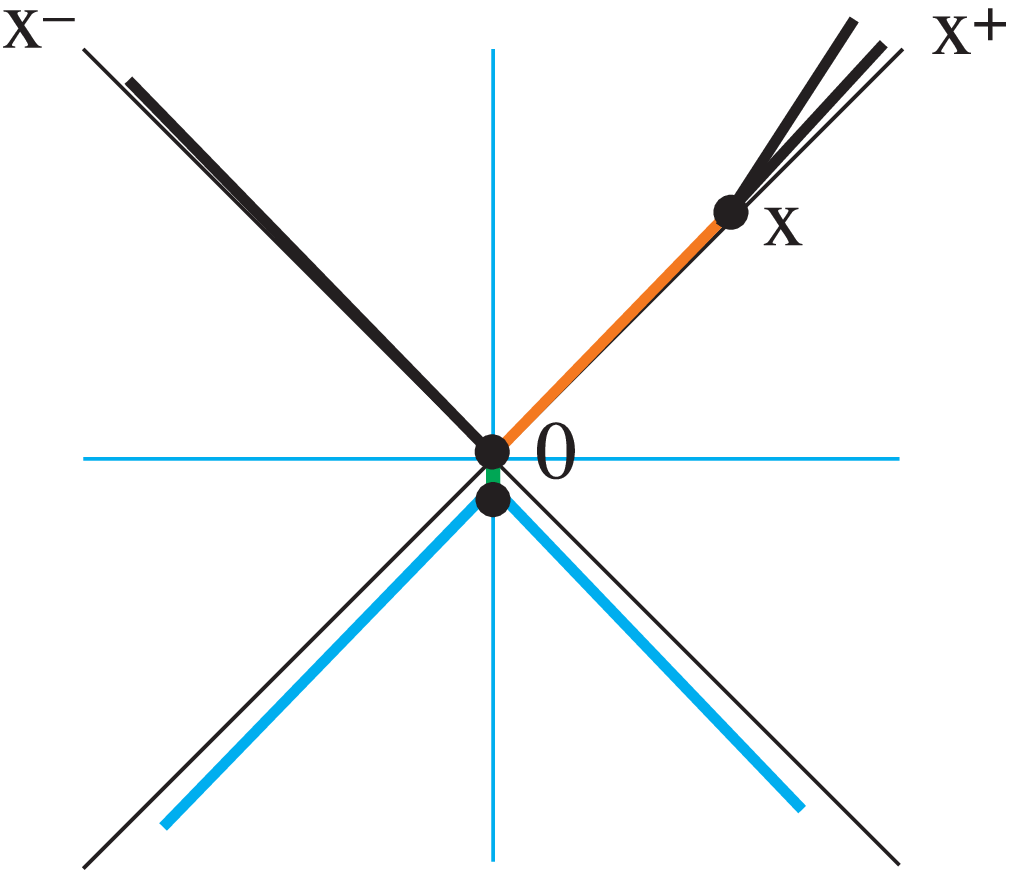 width 6 cm)}
\caption{Correspondence between singularities in momentum space and
the development of the system in space-time.}
\label{nullplane2}
\end{figure}
 
We now return to the singularity structure of $e^+ e^- \to q \bar q
g$. Define $p_1^\mu + p_3^\mu = k^\mu$. Choose null plane coordinates
with $k^+$ large and ${\bf k}_T= {\bf 0}$. Then $k^2 = 2 k^+k^-$
becomes small when
\begin{equation}
k^- = {{\bf p}_{3,T}^2 \over 2p_1^+} + {{\bf p}_{3,T}^2 \over 2p_3^+}
\end{equation}
becomes small. This happens when ${\bf p}_{3,T}$ becomes small with
fixed $p_1^+$ and $p_3^+$, so that the gluon momentum is nearly
collinear with the quark momentum. It also happens when
${\bf p}_{3,T}$ and $p_3^+$ both become small with $p_3^+ \propto
|{\bf p}_{3,T}|$, so that the gluon momentum is soft. ( It also
happens when the quark becomes soft, but there is a
numerator factor that cancels the soft quark singularity.) Thus the
singularities for a soft or collinear gluon correspond to small
$k^-$. 

Now consider the Fourier transform to coordinate space. The quark
propagator in Fig.~\ref{nullplane2} is
\begin{equation}
S_F(k) = \int dx^+ dx^- d{\bf x}\, 
\exp(i[k^+{\blue{x^-}} + k^-{\blue{x^+}} - {\bf k}\cdot{{\bf x}}])\
S_F(x).
\end{equation}
When $k^+$ is large and $k^-$ is small, the contributing values of $x$
have small ${\blue{x^-}}$ and large ${\blue{x^+}}$. Thus the
propagation of the virtual quark can be pictured in space-time as in
Fig.~\ref{nullplane2}. The quark propagates a long distance in
the $x^+$ direction before decaying into a quark-gluon pair. That
is, the singularities that can lead to divergent perturbative cross
sections arise from interactions that happen a long time after the
creation of the initial quark-antiquark pair.

\subsection{Nature of the long-time physics}

\begin{figure}[htb]
\centerline{\DESepsf(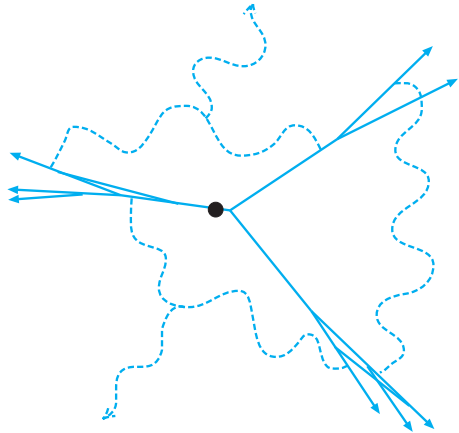 width 6 cm)}
\caption{Typical paths of partons in space contributing to $e^+e^-
\to hadrons$, as suggested by the singularities of perturbative
diagrams. Short wavelength fields are represented by classical paths
of particles.  Long wavelength fields are represented by wavy lines.}
\label{spacetime}
\end{figure}

Imagine dividing the contributions to a scattering cross section into
long-time contributions and short-time contributions. In the
long-time contributions, perturbation theory is out of control, as
indicated in Eq.~(\ref{outofcontrol}). Nevertheless the generic
structure of the long-time contribution is of great interest. This
structure is illustrated in Fig.~\ref{spacetime}. Perturbative
diagrams have big contributions from space-time histories in which
partons move in collinear groups and additional partons are soft and
communicate over large distances, while carrying small momentum.

The picture of Fig.~\ref{spacetime} is suggested by the singularity
structure of diagrams at any fixed order of perturbation theory. Of
course, there could be nonperturbative effects that would
invalidate the picture. Since nonperturbative effects can be
invisible in perturbation theory, one cannot claim that the
structure of the final state indicated in Fig.~\ref{spacetime} is
known to be a consequence of QCD. One can point, however, to some
cases in which one can go beyond fixed order perturbation theory and
sum the most important effects of diagrams of all orders
(for example, Ref.~\citenum{CSbacktoback}). In such cases, the
general picture suggested by Fig.~\ref{spacetime} remains intact. 

We thus find that perturbative QCD suggests a certain structure of
the final state produced in $e^+e^- \to hadrons$: {\blue{the final
state should consist of jets of nearly collinear particles plus
soft particles moving in random directions.}} In fact, this
qualitative prediction is a qualitative success.

Given some degree of qualitative success, we may be bolder and ask
whether perturbative QCD permits quantitative predictions. If we want
quantitative predictions, we will somehow have to find things to
measure that are not sensitive to interactions that happen long after
the basic hard interaction. This is the subject of the next section.

\subsection{The long-time problem}
\label{sec:irsafe}

We have seen that perturbation theory is not effective for long-time
physics. But the detector is a long distance away from the
interaction, so it would seem that long-time physics has to
be present. 

Fortunately, there are some measurements that are {\it not} sensitive
to long-time physics. An example is the total cross section to produce
hadrons in $e^+e^-$ annihilation. Here effects from times $\Delta t
\gg 1/\sqrt s$ cancel because of unitarity. To see why, note that
the quark state is created from the vacuum by a current operator $J$
at some time $t$; it then develops from time $t$ to time $\infty$
according to the interaction picture evolution operator $U(\infty,t)$,
when it becomes the final state $|N\rangle$. The cross section is
proportional to the sum over $N$ of this amplitude times a similar
complex conjugate amplitude with $t$ replaced by a different time
$t^\prime$. We Fourier transform this with $\exp(-i \sqrt
s\,(t-t^\prime))$, so that we can take $\Delta t \equiv t - t^\prime$
to be of order $1/\sqrt s$. Now replacing $\sum |N \rangle\langle N|$
by the unit operator and using the unitarity of the evolution
operators $U$, we obtain
\begin{eqnarray}
\lefteqn{\sum_N \langle 0|J(t^\prime)U(t^\prime,\infty)
|N \rangle\langle N|
U(\infty,t)
J(t)|0\rangle}
\\
&&=
\langle 0|J(t^\prime)U(t^\prime,\infty)
U(\infty,t)
J(t)|0\rangle
=
\langle 0|J(t^\prime)U(t^\prime,t)J(t)|0\rangle.
\nonumber
\end{eqnarray}
Because of unitarity, the long-time evolution has canceled out of
the cross section, and we have only evolution from $t$ to $t^\prime$.

\begin{figure}[htb]
\centerline{\DESepsf(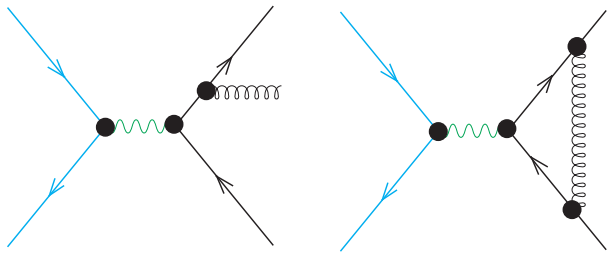 width 8 cm)}
\caption{Cancellation between real and virtual gluon graphs. If we
integrate the real gluon graph on the left times the complex conjugate
of the similar graph with the gluon attached to the antiquark, we will
get an infrared infinity. However the virtual gluon graph on the right
times the complex conjugate of the Born graph is also divergent, as
is the Born graph times the complex conjugate of the virtual gluon
graph. Adding everything together, the infrared infinities cancel.}
\label{eetojetsF}
\end{figure}

There are three ways to view this result. First, we have the formal
argument given above. Second, we have the intuitive understanding
that after the initial quarks and gluons are created in a time
$\Delta t$ of order $1/\sqrt s$, {\it something} will happen with
probability 1. Exactly what happens is long-time physics, but we
don't care about it since we sum over all the possibilities
$|N\rangle$. Third, we can calculate at some finite order of
perturbation theory. Then we see infrared infinities at various
stages of the calculations, but we find that the infinities cancel
between  real gluon emission graphs and virtual gluon graphs. An
example is shown in Fig.~\ref{eetojetsF}.

We see that the total cross section if free of sensitivity to
long-time physics. If the total cross section were all you could look
at, QCD physics would be a little boring. Fortunately, there are
other quantities that are not sensitive to infrared effects. They are
called {\red{infrared safe quantities}}.

To formulate the concept of infrared safety, consider a measured
quantity that is constructed from the cross sections,
\begin{equation}
{d \sigma[n] \over d \Omega_2 d E_3 d \Omega_3
\cdots d E_n d \Omega_n},
\end{equation}
to make $n$ hadrons in $e^+e^-$ annihilation. Here $E_j$ is the
energy of the $j$th hadron and $\Omega_j = (\theta_j,\phi_j)$
describes its direction. We treat the hadrons as effectively massless
and do not distinguish the hadron flavors. Following the notation of
Ref.~\citenum{KS}, let us specify functions ${\cal S}_n$ that describe
the measurement we want, so that the measured quantity is 
\begin{eqnarray}
{\cal I} &=&
{1 \over 2!} \int d \Omega_2\
{d \sigma[2] \over d \Omega_2}\
{\blue{{\cal S}_2}}(p_1^\mu,p_2^\mu)
\nonumber\\
&& +
{1 \over 3!} \int d \Omega_2 d E_3 d \Omega_3\
{d \sigma[3] \over d \Omega_2 d E_3 d \Omega_3}\
{\blue{{\cal S}_3}}(p_1^\mu,p_2^\mu,p_3^\mu)
\nonumber\\
&& +
{1 \over 4!} \int d \Omega_2 d E_3 d \Omega_3 d E_4 d \Omega_4\
\nonumber\\
&&\hskip 2 cm \times
{d \sigma[4] \over d \Omega_2 d E_3 d \Omega_3 d E_4 d \Omega_4}\
{\blue{{\cal S}_4}}(p_1^\mu,p_2^\mu,p_3^\mu,p_4^\mu)
\nonumber\\
&&
+ \cdots .
\end{eqnarray}
The functions $\cal S$ are symmetric functions of their arguments.  In
order for our measurement to be infrared safe, we need
\begin{equation}
{\cal S}_{n+1}
(p_1^\mu,\dots,{\green{(1 - \lambda)}}{\red {p_n^\mu}},{\green{\lambda}} 
{\red{p_n^\mu}}) = 
{\cal S}_n (p_1^\mu,\dots,{\red{p_n^\mu}})
\end{equation}
for  $0\le {\green{\lambda}} \le 1$.

\begin{figure}[htb]
\centerline{\DESepsf(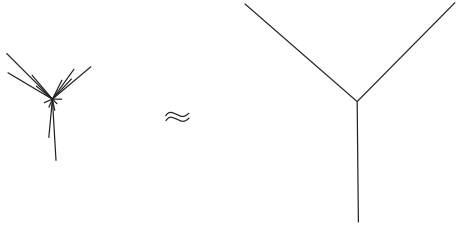 width 6 cm)}
\caption{Infrared safety. In an infrared safe measurement, the three
jet event shown on the left should be (approximately)
equivalent to an ideal three jet event shown on the right.}
\label{irsafe}
\end{figure}

What does this mean?  The physical meaning is that the functions
${\cal S}_n$ and ${\cal S}_{n-1}$ are related in such a way that the
cross section is not sensitive to whether or not a mother particle
divides into two collinear daughter particles that share its
momentum. The cross section is also not sensitive to whether or not a
mother particle decays to a daughter particle carrying all of its
momentum and a soft daughter particle carrying no momentum. The cross
section is also not sensitive to whether or not two collinear
particles combine, or a soft particle is absorbed by a fast particle.
All of these decay and recombination processes can happen with large
probability in the final state long after the hard interaction. But,
by construction, they don't matter as long as the sum of the
probabilities for something to happen or not to happen is one.

Another version of the {\blue{physical meaning}} is that for an
IR-safe quantity a physical event with hadron jets should give
approximately the same measurement as a parton event with each jet
replaced by a parton, as illustrated in Fig.~\ref{irsafe}. To see
this, we simply have to delete soft particles and combine collinear
particles until three jets have become three particles.

In a calculation of the measured quantity $\cal I$, we simply
calculate with partons instead of hadrons in the final state. The
{\blue{calculational meaning}} of the infrared safety condition is
that the infrared infinities cancel. The argument is that the
infinities arise from soft and collinear configurations of the
partons, that these configurations involve long times, and that the
time evolution operator is unitary.

I have started with an abstract formulation of infrared safety. It
would be good to have a few examples. The easiest is the total cross
section, for which
\begin{equation}
{\cal S}_n(p_1^\mu,\dots,p_n^\mu) = 1.
\end{equation}
A less trivial example is the {\blue{thrust}} distribution. One
defines the thrust ${\cal T}_n$ of an $n$ particle event as
\begin{equation}
{\cal T}_n(p_1^\mu,\dots,p_n^\mu)
= \max_{\vec u}
{\sum_{i=1}^n | \vec p_i \cdot \vec u |
\over
\sum_{i=1}^n | \vec p_i |}\ .
\end{equation}
Here $\vec u$ is a unit vector, which we vary to maximize the sum of
the absolute values of the projections of $\vec p_i$ on $\vec u$.
Then the thrust distribution $(1/\sigma_{tot})\,d \sigma/ d T$ is
defined by taking
\begin{equation}
{\cal S}_n(p_1^\mu,\dots,p_n^\mu) =
(1/\sigma_{tot})\
\delta\!\left(T - {\cal T}_n(p_1^\mu,\dots,p_n^\mu)\right)\ .
\end{equation}
It is a simple exercise to show that the thrust of an event is not
affected by collinear parton splitting or by zero momentum partons.
Therefore the thrust distribution is infrared safe.

Another infrared safe quantity is the cross sections to make
{\blue{$n$ jets}}. Here one has to define what one means by a jet.
The definitions used in electron-positron annihilation typically
involve successively combining particles that are nearly collinear
to make the jets. A description can be found in Ref.~\citenum{BKSS}.
I discuss jet cross sections for hadron collisions in
Sec.~\ref{jetproduction}.
 
A final example is the {\blue{ energy-energy correlation}} function
\cite{BBEL}, which measures the average of the product of the energy
in one calorimeter cell times the energy in another calorimeter
cell. One looks at this average as a function of the angular
separation of the calorimeter cells.

\begin{figure}[htb]
\centerline{\DESepsf(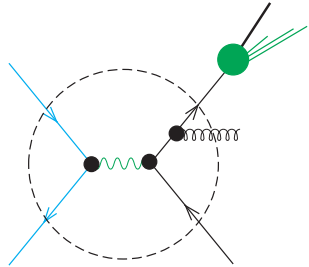 width 4 cm)}
\caption{The cross section for $e^+e^- \to \pi + X$ can be written
as a convolution of a short distance cross section (inside the
dotted line) and a parton decay function.}
\label{decay}
\end{figure}

Before leaving this subject, I should mention another way to 
eliminate sensitivity to long-time physics. Consider the cross
section
\begin{equation}
{ d \sigma(e^+e^- \to \pi + X) \over d E_\pi}.
\end{equation}
This cross section can be written as a convolution of two factors, as
illustrated in Fig.~\ref{decay}. The first factor is a calculated
``hard scattering cross section'' for  $e^+e^- \to {quark} + X$
or $e^+e^- \to {gluon} + X$. The second factor is a ``parton
decay function'' for  ${quark} \to \pi + X$ or ${gluon} \to
\pi + X$. These functions contain the long-time sensitivity and
are to be measured, since they cannot be calculated perturbatively.
However, once they are measured in one process, they can be used for
another process. This final state factorization is similar to the
initial state factorization involving parton distribution functions,
which we will discuss later. (See
Refs.~\citenum{handbook},\citenum{ESW},\citenum{CSparton} for more
information.)

\section{ The smallest time scales}
\label{smallest}

In this section, I explore the physics of time scales smaller than
$1/\sqrt s$. One way of looking at this physics is to say that it is
plagued by infinities and we can manage to hide the infinities. A
better view is that the short-time physics contains
wonderful truths that we would like to discover -- truths about
grand unified theories, quantum gravity and the like. However,
quantum field theory is arranged so as to effectively hide the truth
from our experimental apparatus, which can probe with a time
resolution of only an inverse half TeV.

I first outline what renormalization does to hide the ugly
infinities or the beautiful truth. Then I describe how
renormalization leads to the running coupling. Because of
renormalization, calculated quantities depend on a
renormalization scale. I look at how this dependence works and how
the scale can be chosen. Finally, I discuss how one can use experiment
to look for the hidden physics beyond the Standard Model, taking
high $E_T$ jet production in hadron collisions as an example.

\subsection{What renormalization does}

\begin{figure}[htb]
\centerline{\DESepsf(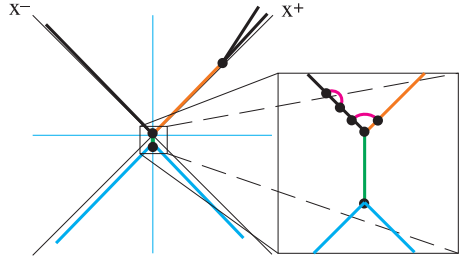 width 6 cm)}
\caption{Renormalization. The effect of the very small time
interactions pictured are absorbed into the running coupling.}
\label{nullplane3}
\end{figure}

In any Feynman graph, one can insert perturbative corrections to the
vertices and the propagation of particles, as illustrated in 
Fig.~\ref{nullplane3}. The loop integrals in these graphs will get big
contributions from momenta much larger than $\sqrt s$. That is, there
are big contributions from interactions that happen on time scales
much smaller than $1/\sqrt s$. I have tried to illustrate this in the
figure. The virtual vector boson propagates for a time $1/\sqrt s$,
while the virtual fluctuations that correct the electroweak vertex
and the quark propagator occur over a time $\Delta t$ that can be
much smaller than $1/\sqrt s$. 

Let us pick an ultraviolet cutoff $M$ that is much larger than $\sqrt
s$, so that we calculate the effect of fluctuations with $1/M <
\Delta t$ exactly, up to some order of perturbation theory. What,
then, is the effect of virtual fluctuations on smaller time scales,
$\Delta t$ with $\Delta t< 1/M$ but, say, $\Delta t$ still larger
than $t_{\rm Plank}$, where gravity takes over? Let us suppose that
we are willing to neglect contributions to the cross section that are
of order $\sqrt s /M$ or smaller compared to the cross section
itself. Then there is a remarkable theorem \cite{JCCbook}: the effects
of the fluctuations are not particularly small, but they can be
absorbed into changes in the couplings of the theory. (There are also
changes in the masses of the theory and  adjustments to the
normalizations of the field operators, but we can concentrate on the
effect on the couplings.)

The program of absorbing very short-time physics into a few
parameters goes under the name of renormalization. There are several
schemes available for renormalizing. Each of them involves the
introduction of some scale parameter that is not intrinsic to the
theory but tells how we did the renormalization. Let us agree to use
\MSbar\ renormalization (see Ref.~\citenum{JCCbook} for details). Then
we introduce an  \MSbar\ renormalization scale $\mu$. A good (but
approximate) way of thinking of $\mu$ is that the physics of time
scales $\Delta t \ll 1/\mu$ is removed from the perturbative
calculation. The effect of the small time physics is accounted for by
adjusting the value of the strong coupling, so that its value depends
on the scale that we used: $\alpha_s = \alpha_s(\mu)$. (The value of
the electromagnetic coupling also depends on $\mu$.)

\subsection{The running coupling}

\begin{figure}[htb]
\centerline{\DESepsf(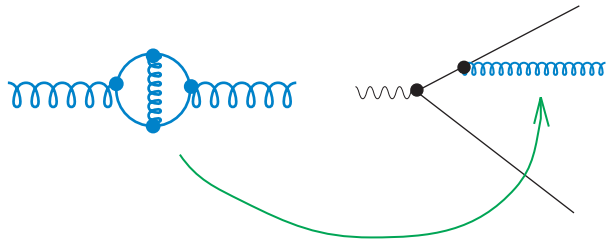 width 8 cm)}
\caption{Short-time fluctuations in the propagation of the gluon
field absorbed into the running strong coupling.}
\label{runninggraph}
\end{figure}

We account for time scales much smaller than $1/\mu$ by using
the running coupling $\alpha_s(\mu)$. That is, a fluctuation such as
that illustrated in Fig.~\ref{runninggraph} can be dropped from a
calculation and absorbed into the running coupling that describes
the probability for the quark in the figure to emit the gluon.
The $\mu$ dependence of $\alpha_s(\mu)$ is given by a certain
differential equation, called the renormalization group equation (see
Ref.~\citenum{JCCbook}):
\begin{equation}
{ d \over d \ln(\mu^2)}\,{ \alpha_s(\mu) \over \pi} = 
\beta(\alpha_s(\mu)) =
- \beta_0\, \left(\alpha_s(\mu)\over \pi\right)^2
- \beta_1\, \left(\alpha_s(\mu)\over \pi\right)^3
+\cdots.
\label{rengrp}
\end{equation}
One calculates the beta function $\beta(\alpha_s)$
perturbatively in QCD. The first coefficient, with the conventions
used here, is
\begin{equation}
\beta_0 = (33 - 2\,N_f)/12\,,
\end{equation}
where $N_f$ is the number of quark flavors.

Of course, at time scales smaller than a very small cutoff $1/M$ (at
the ``GUT scale,'' say) there is completely different physics
operating. Therefore, if we use just QCD to adjust the strong
coupling, we can say that we are accounting for the physics between
times $1/M$ and $1/\mu$. The value of $\alpha_s$ at
$\mu_0 \approx M$ is then the boundary condition for the differential
equation.

\begin{figure}[htb]
\centerline{\DESepsf(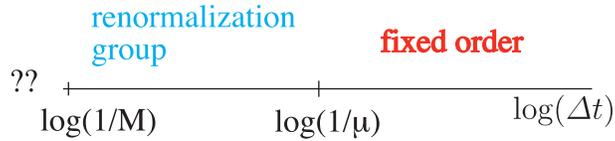 width 8 cm)}
\caption{Distance scales accounted for by explicit fixed order
perturbative calculation and by use of the renormalization group.}
\label{scales.eps}
\end{figure}

The renormalization group equation sums the effects of short-time
fluctuations of the fields. To see what one means by ``sums''
here, consider the result of solving the renormalization
group equation with all of the $\beta_i$ beyond $\beta_0$ set to
zero:
\begin{eqnarray}
\alpha_s(\mu) &\approx&
\alpha_s(M) -  (\beta_0/\pi) \ln(\mu^2/M^2)\ \alpha_s^2(M)
\nonumber\\ 
&& \quad +
 (\beta_0/\pi)^2 \ln^2(\mu^2/M^2)\ \alpha_s^3(M) + \cdots
\nonumber\\
&=&
{\alpha_s(M)\over 1 + (\beta_0/\pi)\,\alpha_s(M) \ln(\mu^2/M^2) }.
\label{running}
\end{eqnarray}
A series in powers of $\alpha_s(M)$ -- that is the strong coupling at
the GUT scale -- is summed into a simple function of $\mu$. Here
$\alpha_s(M)$ appears as a parameter in the solution.

Note a crucial and wonderful fact. The value of $\alpha_s(\mu)$
decreases as $\mu$ increases. This is called ``asymptotic freedom.''
Asymptotic freedom implies that QCD acts like a weakly interacting
theory on short time scales. It is true that quarks and gluons are
strongly bound inside nucleons, but this strong binding is the
result of weak forces acting collectively over a long time.

In Eq.~(\ref{running}), we are invited to think of the graph of
$\alpha_s(\mu)$ versus $\mu$. The differential equation that
determines this graph is characteristic of QCD. There could, however,
be different versions of QCD with the same differential equation
but different curves, corresponding to different boundary values
$\alpha_s(M)$. Thus the parameter $\alpha_s(M)$ tells us which
version of QCD we have. To determine this parameter, we consult
experiment. Actually, Eq.~(\ref{running}) is not the most convenient
way to write the solution for the running coupling. A better
expression is 
\begin{equation}
\alpha_s(\mu) \approx 
{ \pi
\over
\beta_0 \ln(\mu^2/\Lambda^2)}.
\end{equation}
Here we have replaced $\alpha_s(M)$ by a different (but
completely equivalent) parameter $\Lambda$. A third form of the
running coupling is
\begin{equation}
\alpha_s(\mu) \approx 
{\alpha_s(M_Z)\over 1 + (\beta_0/\pi)\,\alpha_s(M_Z)
\ln(\mu^2/M_Z^2) }.
\end{equation}
Here the value of $\alpha_s(\mu)$ at $\mu = M_Z$ labels the version
of QCD that obtains in our world.

In any of the three forms of the running coupling, one should
revise the equations to account for the second term in the beta
function in order to be numerically precise.

\subsection{The choice of scale}

In this section, we consider the choice of the renormalization
scale $\mu$ in a calculated cross section.  Consider, as an
example, the cross section for $e^+ e^-\to {\rm hadrons }$ via
virtual photon decay. Let us write this cross section in the form
\begin{equation}
\sigma_{\rm tot} = {4 \pi \alpha^2\over s}\left(\sum_f Q_f^2 \right)
\left[1 + \Delta \right].
\end{equation}
Here $s$ is the square of the c.m.\ energy, $\alpha$ is
$e^2/(4\pi)$, and $Q_f$ is the electric charge in units of $e$
carried by the quark of flavor $f$, with $f = u,d,s,c,b$. The
nontrivial part of the calculated cross section is the quantity
$\Delta$, which contains the effects of the strong interactions.
Using \MSbar\ renormalization with scale $\mu$, one finds (after a lot
of work) that $\Delta$ is given by \cite{Levan} 
\begin{eqnarray}
&&\Delta  = 
{\alpha_s({\blue{\mu}})\over \pi}
 + \left[1.4092 + 1.9167\ \ln\left({\blue{\mu^2}} / s\right)
\right] \left({\alpha_s({\blue{\mu}})\over
\pi}\right)^{\!2}
\nonumber\\ 
&&\quad + 
\left[ -12.805 + 7.8186\ \ln\left({\blue{\mu^2}} / s\right) 
+ 3.674\ \ln^2\!\left({\blue{\mu^2}} / s\right)\right]
\left({\alpha_s({\blue{\mu}})\over \pi}\right)^{\!3}
\nonumber\\
&&\quad
+\cdots.
\label{eecalc}
\end{eqnarray}
Here, of course, one should use for $\alpha_s(\mu)$ the solution of
the renormalization group equation (\ref{rengrp}) with at least
two terms included.

As discussed in the preceding subsection, when we renormalize
with scale $\mu$, we are defining what we mean by the strong
coupling. Thus $\alpha_s$ in Eq.~(\ref{eecalc}) depends on $\mu$.  The
perturbative coefficients in Eq.~(\ref{eecalc}) also depend on $\mu$.
On the other hand, the physical cross section {\red{does not}} depend
on $\mu$:
\begin{equation}
{d \over d \ln \mu^2}\,\Delta = 0.
\label{nomu}
\end{equation}
That is because $\mu$ is just an artifact of how we organize
perturbation theory, not a parameter of the underlying theory.

Let us consider Eq.~(\ref{nomu}) in more detail. Write $\Delta$ in
the form
\begin{equation}
\Delta \sim \sum_{n=1}^\infty c_n(\mu)\ \alpha_s(\mu)^n.
\end{equation}
If we differentiate not the complete infinite sum but just the first
$N$ terms, we get minus the derivative of the sum from $N+1$ to
infinity. This remainder is of order $\alpha_s^{N+1}$ as $\alpha_s
\to 0$. Thus
\begin{equation}
{d \over d \ln \mu^2}\sum_{n=1}^N c_n(\mu)\ \alpha_s(\mu)^n \sim {\cal
O}(\alpha_s(\mu)^{N+1}).
\end{equation}
That is, the harder we work calculating more terms, the less the
calculated cross section depends on $\mu$.

Since we have not worked infinitely hard, the calculated cross
section depends on $\mu$. What choice shall we make for $\mu$?
Clearly, $\ln\left(\mu^2 / s\right)$ should not be big. Otherwise
the coefficients $c_n(\mu)$ are large and the ``convergence'' of
perturbation theory will be spoiled. There are some who will argue
that one scheme or the other for choosing $\mu$ is the ``best.'' You
are welcome to follow whichever advisor you want. I will show you
below that for a well behaved quantity like $\Delta$ the precise
choice makes little difference, as long as you obey the
common sense prescription that $\ln\left(\mu^2 / s\right)$ not be
big. 

\subsection{An example}
\label{errorest}

\begin{figure}[htb]
\vskip 2.5 cm
\leftline{\hskip 1.8cm $\Delta(\mu)$}
\vskip 2.5 cm
\centerline{\hskip 1cm $\ln_2(\mu/\sqrt s)$}
\vskip -5.5 cm
\centerline{\DESepsf(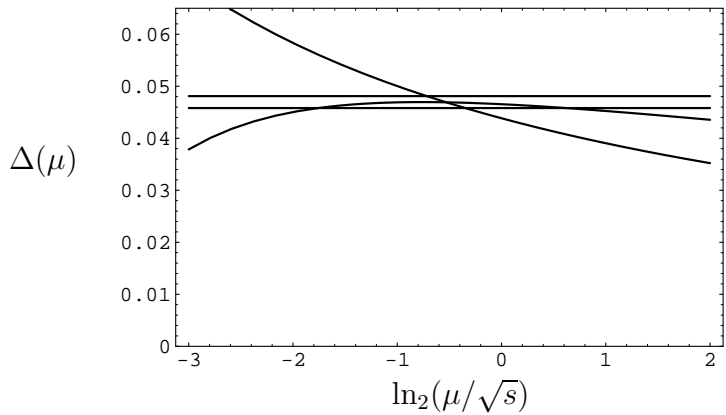 width 8 cm)}
\vskip 0.5 cm
\caption{Dependence of $\Delta(\mu)$ on the \MSbar\ renormalization
scale $\mu$. The falling curve is $\Delta_1$. The flatter curve is
$\Delta_2$. The horizontal lines indicates the amount of variation of
$\Delta_2$ when $\mu$ varies by a factor 2.}
\label{eemuA}
\end{figure}

Let us consider a quantitative example of how $\Delta(\mu)$ depends
on $\mu$. This will also give us a chance to think about the
theoretical error caused by replacing $\Delta$ by the sum
$\Delta_n$ of the first $n$ terms in its perturbative expansion. Of
course, we do not know what this error is. All we can do is provide
an estimate. (Our discussion will be rather primitive. For a more
detailed error estimate for the case of the hadronic width of the $Z$
boson, see Ref.~\citenum{SandS}.)

Let us think of the error estimate in the spirit of a ``$1\,\sigma$''
theoretical error: we would be surprised if $|\Delta_n - \Delta|$ were
much less than the error estimate and we would also be surprised
if this quantity were much more than the error estimate. Here,
one should exercise a little caution. We have no reason to
expect that theory errors are gaussian distributed. Thus a
$4\,\sigma$ difference between $\Delta_n$ and $\Delta$ is not out of
the question, while a $4\,\sigma$ fluctuation in a measured quantity
with purely statistical, gaussian errors {\it is} out of the
question.

Take $\alpha_s(M_Z) = 0.117$, $\sqrt s = 34 \GeV$, 5 flavors. In
Fig.~\ref{eemuA}, I plot $\Delta(\mu)$ versus ${\blue{p}}$ defined by
\begin{equation}
\mu = 2^{\blue{p}} \sqrt s.
\end{equation}
The steeply falling curve is the order $\alpha_s^1$ approximation to
$\Delta(\mu)$,  $\Delta_1(\mu) = {\alpha_s(\mu)/ \pi}$. Notice that
if we change $\mu$ by a factor 2, $\Delta_1(\mu)$ changes by about
0.006. If we had no other information than this, we might pick
$\Delta_1(\sqrt s) \approx 0.044$ as the ``best'' value and assign a
$\pm 0.006$ error to this value. (There is no special magic to the use
of a factor of 2 here. The reader can pick any factor that seems
reasonable.)

Another error estimate can be based on the simple expectation that
the coefficients of $\alpha_s^n$ are of order 1 for the first few
terms. (Eventually, they will grow like $n!$. Ref.~\citenum{SandS}
takes this into account, but we ignore it here.) Then the first
omitted term should be of order $\pm 1 \times \alpha_s^2 \approx
\pm 0.020$ using $\alpha_s(34\ \GeV) \approx 0.14$. Since this is
bigger than the previous $\pm 0.006$ error estimate, we keep this
larger estimate: $\Delta \approx 0.044 \pm 0.020$.

Returning now to Fig.~\ref{eemuA}, the second curve is the order
$\alpha_s^2$ approximation, $\Delta_2(\mu)$. Note that $\Delta_2(\mu)$
is less dependent on $\mu$ than $\Delta_1(\mu)$.

What value would we now take as our best estimate of $\Delta$? One
idea is to choose the value of $\mu$ at which $\Delta_2(\mu)$ is
least sensitive to $\mu$. This idea is called the {\it principle of
minimal sensitivity} \cite{PMS}:
\begin{equation}
\Delta_{PMS} = \Delta(\mu_{PMS})\,, \hskip 1 cm 
\left[{d \Delta(\mu) \over d \ln \mu}\right]_{\mu = \mu_{PMS}} = 0.
\end{equation}
This prescription gives $\Delta \approx 0.0470$. Note that this is
about 0.003 away from our previous estimate, $\Delta \approx 0.0440$.
Thus our previous error estimate of 0.020 was too big, and we should
be surprised that the result changed so little. We can make a new
error estimate by noting that $\Delta_2(\mu)$ varies by about 0.0012
when
$\mu$ changes by a factor $2$ from $\mu_{PMS}$. Thus we might
estimate that $\Delta \approx 0.0470$ with an error of $\pm 0.0012$.
This estimate is represented by the two horizontal lines in 
Fig.~\ref{eemuA}.

An alternative error estimate can be based on the next term being of
order $\pm 1 \times \alpha_s^3(34\ GeV) \approx 0.003$. Since this is
bigger than the previous $\pm 0.0012$ error estimate, we keep this
larger estimate: $\Delta \approx 0.0470 \pm 0.003$.

I should emphasize that there are other ways to pick the ``best''
value for $\Delta$. For instance, one can use the BLM method
\cite{BLM}, which is based on choosing the $\mu$ that sets to zero
the coefficient of the number of quark flavors in $\Delta_2(\mu)$.
Since the graph of $\Delta_2(\mu)$ is quite flat, it makes very
little difference which method one uses.

\begin{figure}[htb]
\vskip 2.5 cm
\leftline{\hskip 1.8cm $\Delta(\mu)$}
\vskip 2.5 cm
\centerline{\hskip 1.0cm$\log_2(\mu/\sqrt s)$}
\vskip -5.5 cm
\centerline{\DESepsf(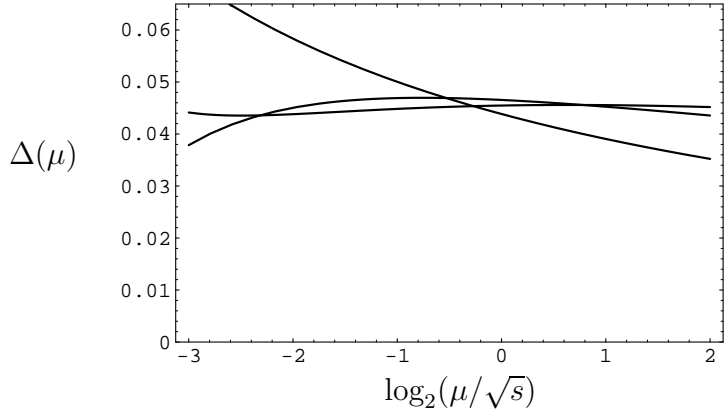 width 8 cm)}
\vskip 0.5 cm
\caption{Dependence of $\Delta(\mu)$ on the \MSbar\ renormalization
scale $\mu$. The falling curve is $\Delta_1$. The flatter curve is
$\Delta_2$. The still flatter curve is $\Delta_3$.}
\label{eemuB}
\end{figure}

Now let us look at $\Delta(\mu)$ evaluated at order $\alpha_s^3$,
$\Delta_3(\mu)$. Here we make use of the full formula in
Eq.~(\ref{eecalc}). In Fig.~\ref{eemuB}, I plot $\Delta_3(\mu)$ along
with $\Delta_2(\mu)$ and $\Delta_1(\mu)$. The variation of
$\Delta_3(\mu)$ with $\mu$ is smaller than that of
$\Delta_2(\mu)$. The improvement is not overwhelming, but is
apparent particularly at small $\mu$.

\begin{figure}[htb]
\vskip 2.5 cm
\leftline{\hskip 1.8cm $\Delta(\mu)$}
\vskip 2 cm
\centerline{\hskip 1.0cm $\log_2(\mu/\sqrt s)$}
\vskip -5.5 cm
\centerline{\DESepsf(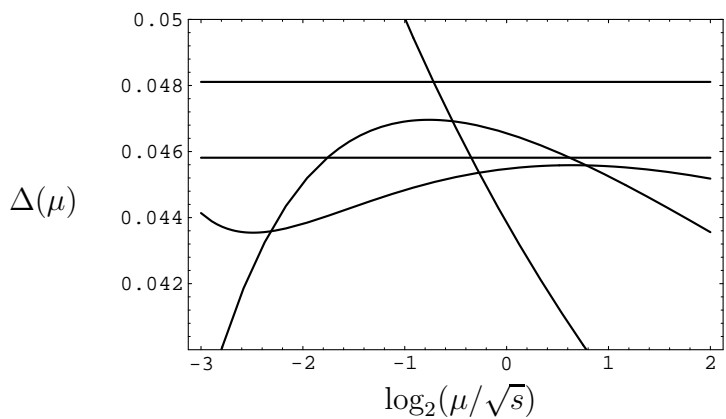 width 8 cm)}
\vskip 0.5 cm
\caption{Dependence of $\Delta(\mu)$ on the \MSbar\ renormalization
scale $\mu$ with an expanded scale. The falling curve is $\Delta_1$.
The flatter curve is $\Delta_2$. The still flatter curve is
$\Delta_3$. The horizontal lines represent the variation of
$\Delta_2$ when $\mu$ varies by a factor 2.}
\label{eemuC}
\end{figure}

It is a little difficult to see what is happening in
Fig.~\ref{eemuB}, so I show the same thing with an expanded scale in
Fig.~\ref{eemuC}. (Here the error band based on the $\mu$
dependence of $\Delta_2$ is also indicated. Recall that we decided
that this error band was an underestimate.) The curve for
$\Delta_3(\mu)$ has zero derivative at two places. The corresponding
values are $\Delta \approx 0.0436$ and $\Delta \approx 0.0456$. If I
take the best value of $\Delta$ to be the average of these two values
and the error to be half the difference, I get $\Delta \approx 0.0446
\pm 0.0010$.

The alternative error estimate is $\pm 1 \times \alpha_s^4(34\ \GeV)
\approx 0.0004$. We keep the larger error estimate of $\pm 0.0010$.

Was the previous error estimate valid? We guessed 
$\Delta \approx 0.0470 \pm 0.003$. Our new best estimate is 0.0446.
The difference is 0.0024, which is in line with our previous error
estimate. Had we used the error estimate $\pm 0.0012$ based on the
$\mu$ dependence, we would have underestimated the difference,
although we would not have been too far off.

\subsection{Beyond the Standard Model}

We have seen how the renormalization group enables us to account for
QCD physics at time scales much smaller than $\sqrt s$, as indicated
in Fig.~\ref{scales}.  However, at some scale $\Delta t \sim 1/M$, we
run into the unknown!

{\blue{How can we see the unknown}} in current experiments?
First, the unknown physics affects $\alpha_s$, $\alpha_{em}$,
$\sin^2(\theta_W)$. Second, the unknown physics affects masses of
$u,d,\dots,e,\mu,\dots$. That is, the unknown physics (presumably)
determines the {\blue{parameters}} of the Standard Model. These
parameters have been well measured. Thus, a Nobel prize awaits the
physicist who figures out how to use a model for the unknown physics
to predict these parameters.

\begin{figure}[htb]
\centerline{\DESepsf(scales.eps width 8 cm)}
\caption{Time scales accounted for by fixed order perturbative
calculations and by use of the renormalization group.}
\label{scales}
\end{figure}

\begin{figure}[htb]
\centerline{\DESepsf(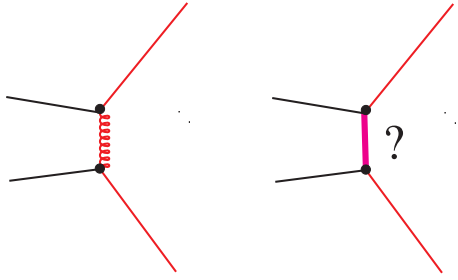 width 6 cm)}
\caption{New physics at a TeV scale. In the first diagram, quarks
scatter by gluon exchange. In the second diagram, the quarks
exchange a new object with a TeV mass, or perhaps exchange some of
the constituents out of which quarks are made.}
\label{newphys}
\end{figure}

There is another way that as yet unknown physics can affect
current experiments. Suppose that quarks can scatter by the exchange
of some new particle with a heavy mass $M$, as illustrated in
Fig.~\ref{newphys}, and suppose that this mass is not too enormous,
only a few TeV. Perhaps the new particle isn't a particle at all, but
is a pair of constituents that live inside of quarks. As mentioned
above, this physics affects the parameters of the Standard Model.
However, unless we can predict the parameters of the Standard Model,
this effect does not help us. There is, however, another possible
clue. The physics at the TeV scale can introduce {\blue{new terms}}
into the lagrangian that we can investigate in current experiments.

In the second diagram in Fig.~\ref{newphys}, the two vertices are
never at a separation in time greater than $1/M$, so that our low
energy probes cannot resolve the details of the structure. As long
as we stick to low energy probes, $\sqrt s \ll M$, the effect of the
new physics can be summarized by adding new terms to the lagrangian of
QCD. A typical term might be
\begin{equation}
{\blue{\Delta{\cal L} = {\tilde g^2 \over M^2} \
\bar \psi \gamma^\mu \psi\
\bar \psi \gamma_\mu \psi.}}
\label{newterm}
\end{equation}
There is a factor $\tilde g^2$ that represents how well the new
physics couples to quarks. The most important factor is the factor
$1/M^2$. This factor must be there: the product of field operators has
dimension 6 and the lagrangian has dimension 4, so there must be a
factor with dimension $-2$. Taking this argument one step further,
the product of field operators in $\Delta {\cal L}$ must have a
dimension greater than 4 because any product of field
operators having dimension equal to or less than 4 that respects the
symmetries of the Standard Model is already included in the
lagrangian of the Standard Model.

\subsection{Looking for new terms in the effective lagrangian}

How can one detect the presence in the lagrangian of a term like that
in Eq.~(\ref{newterm})? These terms are small. Therefore we need
either a high precision experiment, or an experiment that looks for
some effect that is forbidden in the Standard Model, or an experiment
that has moderate precision and operates at energies that are as high
as possible.

Let us consider an example of the last of these
possibilities, $p + \bar p \to jet + X$ as a function of the
transverse energy ($\sim P_T$) of the jet. The new term in the
lagrangian should add a little bit to the observed cross section that
is not included in the standard QCD theory. When the transverse
energy $E_T$ of the jet is small compared to $M$, we expect 
\begin{equation}
{\sienna{{{\rm Data} - {\rm Theory} \over {\rm Theory}}}}
\propto \tilde g^2 { {\red{E_T^2}} \over M^2}.
\label{dataup}
\end{equation}
Here the factor $\tilde g^2 / M^2$ follows because $\Delta {\cal L}$
contains this factor. The factor $E_T^2$ follows because the left
hand side is dimensionless and $E_T$ is the only factor with
dimension of mass that is available.

\begin{figure}[htb]
\centerline{\DESepsf(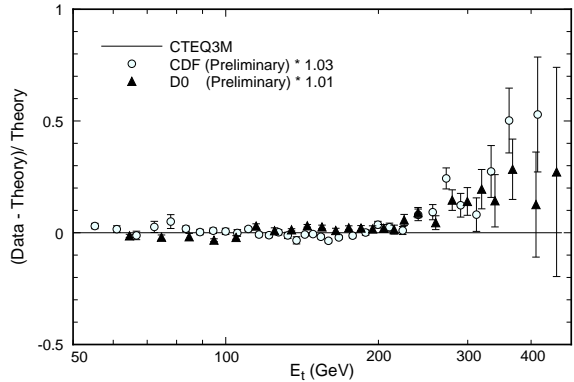 width 8 cm)}
\caption{Jet cross sections from CDF and D0 compared to QCD theory.
(Data $-$ Theory)/Theory is plotted versus the transverse
energy $E_T$ of the jet. The theory here is next-to-leading order
QCD using the CTEQ3M parton distribution. Source:
Ref.~\protect\citenum{CTEQ4}}
\label{Jetcteq3}
\end{figure}

In Fig.~\ref{Jetcteq3}, I show a plot comparing experimental jet
cross sections from CDF \cite{CDFjet} and D0 \cite{D0jet} compared to
next-to-leading order QCD theory. The theory works fine for $E_T< 200
\GeV$, but for $200 \GeV < E_T$, there appears to be a systematic
deviation of just the form anticipated in Eq.~(\ref{dataup}). 

This example illustrates the idea of how small distance physics
beyond the Standard Model can leave a trace in the form of small
additional terms in the effective lagrangian that controls physics
at currently available energies. However, in this case, there is some
indication that the observed effect might be explained by some
combination of the experimental systematic error and the
uncertainties inherent in the theoretical prediction \cite{CTEQjet}.
In particular, the prediction is sensitive to the distributions of
quarks and gluons contained in the colliding protons, and the gluon
distribution in the kinematic range of interest here is rather poorly
known. In the next section, we turn to the definition, use, and
measurement of the distributions of quarks and gluons in hadrons.

\section{Deeply inelastic scattering}

Until now, I have concentrated on hard scattering processes with
leptons in the initial state. For such processes, we have seen that
the hard part of the process can be described using perturbation
theory because $\alpha_s(\mu)$ gets small as $\mu$ gets large.
Furthermore, we have seen how to isolate the hard part of the
interaction by choosing an infrared safe observable. But what about
hard processes in which there are hadrons in the initial state? Since
the fundamental hard interactions involve quarks and gluons,
the theoretical description necessarily involves a description of how
the quarks and gluons are distributed in a hadron. Unfortunately,
the distribution of quarks and gluons in a hadron is controlled by
long-time physics. We cannot calculate the relevant
distribution functions perturbatively (although a calculation in
lattice QCD might give them, in principle). Thus we must find how to
separate the short-time physics from the parton distribution
functions and we must learn how the parton distribution functions
can be determined from the experimental measurements.

In this section, I discuss parton distribution functions and their
role in deeply inelastic lepton scattering (DIS). This includes  $e +
p \to e + X$ and $\nu + p \to e + X$ where the momentum transfer
from the lepton is large. I first outline the kinematics of deeply
inelastic scattering and define the structure functions $F_1$, $F_2$
and $F_3$ used to describe the process. By examining the
space-time structure of DIS, we will see how the cross section can
be written as a convolution of two factors, one of which is the
parton distribution functions and the other of which is a cross
section for the lepton to scatter from a quark or gluon. This
factorization involves a scale $\mu_F$ that, roughly speaking,
divides the soft from the hard regime; I discuss the dependence of
the calculated cross section on $\mu_F$. With this groundwork laid,
I give the  \MSbar\ definition of parton distribution functions in
terms of field operators and discuss the evolution equation for the
parton distributions. I close the section with some comments on how
the parton distributions are, in practice, determined from experiment.

\subsection{Kinematics of deeply inelastic lepton scattering}

\begin{figure}[htb]
\centerline{\DESepsf(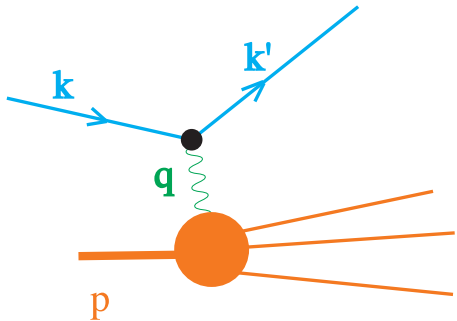 width 6 cm)}
\caption{Kinematics of deeply inelastic scattering}
\label{DISA}
\end{figure}

In deeply inelastic scattering, a lepton with momentum $k^\mu$
scatters on a hadron with momentum $p^\mu$. In the
final state, one observes the scattered lepton with momentum
$k^{\prime\mu}$ as illustrated in Fig.~\ref{DISA}. The momentum
transfer
\begin{equation}
q^\mu = k^\mu - k^{\prime \mu}
\end{equation}
is carried on a photon, or a $W$ or $Z$ boson.

The interaction between the vector boson and the hadron depends
on the variables $q^\mu$ and $p^\mu$. From these two vectors we can
build two scalars (not counting $m^2 = p^2$). The first variable is
\begin{equation}
{\blue{Q^2}} = - q^2 ,
\end{equation}
where the minus sign is included so that $Q^2$ is positive. The
second scalar is the dimensionless Bjorken variable, 
\begin{equation}
{\blue{x_{\rm bj}}} = { Q^2 \over 2 p \cdot q}.
\end{equation}
(In the case of scattering from a nucleus containing $A$ nucleons,
one replaces $p^\mu$ by $p^\mu /A$ and defines $x_{\rm bj} = A\,
{Q^2 / (2 p \cdot q)}$.)

One calls the scattering {\it deeply inelastic} if $Q^2$ is large
compared to $1\ \GeV^2$. Traditionally, one speaks of the {\it
scaling limit}, $Q^2 \to \infty$ with $x_{\rm bj}$ fixed. Actually,
the asymptotic theory to be described below works pretty well if
$Q^2$ is bigger than, say, $4\, \GeV^2$ and $x_{\rm bj}$ is anywhere
in the experimentally accessible range, roughly $10^{-4} < x_{\rm bj}
< 0.5$.

The invariant mass squared of the hadronic final state is
$W^2 = (p + q)^2$. In the scaling regime of large $Q^2$ one has
\begin{equation}
W^2 = m^2 + {1-x_{\rm bj} \over x_{\rm bj}}\, Q^2 \gg m^2.
\end{equation}
This justifies saying that the scattering is not only inelastic but
deeply inelastic.

We have spoken of the scalar variables that one can form from
$p^\mu$ and $q^\mu$.  Using the lepton momentum $k^\mu$, one can
also form the dimensionless variable
\begin{equation}
y = {p\cdot q \over p\cdot k}.
\end{equation}

\subsection{Structure functions for DIS}

One can make quite a lot of progress in understanding the theory of
deeply inelastic scattering without knowing anything about QCD
except its symmetries.  One expresses the cross section in terms of
three structure functions, which are functions of $x_{\rm bj}$ and
$Q^2$ only.

Suppose that the initial lepton is a neutrino, $\nu_\mu$, and the
final lepton is a muon. Then in Fig.~\ref{DISA} the exchanged vector
boson, call it $V$,  is a $W$ boson, with mass $M_V = M_W$.
Alternatively, suppose that both the initial and final leptons are
electrons and let the exchanged vector boson be a photon, with mass
$M_V = 0$. This was the situation in the original DIS experiments at
SLAC in the late 1960's. In experiments with sufficiently large
$Q^2$, $Z$ boson exchange should be considered along with photon
exchange, and the formalism described below must be augmented.

Given only the electroweak theory to tell us how the vector boson
couples to the lepton, one can write the cross section in the form
\begin{equation}
d \sigma = {4 \alpha^2 \over s}{d^3 {\bf k}^\prime \over 2|{\bf k}^\prime|}
{C_V \over (q^2 - M_V^2)^2}\,
{\green{L^{\mu\nu}}}(k,q)\,{\blue{W_{\mu\nu}}}(p,q),
\label{DIS1}
\end{equation}
where $C_V$ is 1 in the case that $V$ is a photon and $1/(64
\,\sin^4\theta_W)$ in the case that $V$ is a $W$ boson.
The tensor ${\green{L^{\mu\nu}}}$ describes the lepton coupling to the
vector boson and has the form
\begin{equation}
{\green{L^{\mu\nu}}} = {1 \over 2} 
{\rm Tr}\left(k\cdot \gamma\ \gamma^\mu k^\prime\cdot
\gamma\ \gamma^\nu 
\right).
\label{DIS2}
\end{equation}
in the case that $V$ is a photon. For a $W$ boson, one has
\begin{equation}
{\green{L^{\mu\nu}}} =  
{\rm Tr}\left(k\cdot \gamma\ \Gamma^\mu k^\prime\cdot
\gamma\ \Gamma^\nu 
\right).
\label{DIS3}
\end{equation}
where $\Gamma^\mu$ is  $\gamma^\mu(1-\gamma_5)$ for a $W^+$
boson ($\nu \to W^+  \ell$) or $\gamma^\mu(1+\gamma_5)$ for a $W^-$
boson ($\bar \nu \to W^-  \bar\ell$). See Ref.~\citenum{handbook}.

The tensor $W^{\mu\nu}$ describes the coupling of the vector boson
to the hadronic system. It depends on $p^\mu$ and $q^\mu$. We know
that it is Lorentz invariant and that $W^{\nu\mu} = W^{\mu\nu*}$. We
also know that the current to which the vector boson couples is
conserved (or in the case of the axial current, conserved in the
absence of quark masses, which we here neglect) so that $q_\mu
W^{\mu\nu} = 0$. Using these properties, one finds three possible
tensor structures for $W^{\mu\nu}$. Each of the three tensors
multiplies a structure function, $F_1$, $F_2$ or $F_3$, which, since
it is a Lorentz scalar, can depend only on the invariants $x_{\rm
bj}$ and $Q^2$. Thus
\begin{eqnarray}
{\blue{W_{\mu\nu}}} &=&
-\left(
g_{\mu\nu}- {q_\mu q_\nu \over q^2}
\right)
 {\red{F_1(x_{\rm bj},Q^2)}}
\nonumber\\ 
&&\quad 
+ 
\left(
p_\mu - q_\mu {p\cdot q \over q^2}
\right)
\left(
p_\nu - q_\nu {p\cdot q \over q^2}
\right)
{ 1\over p \cdot q }\
 {\red{F_2(x_{\rm bj},Q^2)}}
\nonumber\\ 
&&\quad
-i \epsilon_{\mu\nu\lambda\sigma}p^\lambda q^\sigma
{ 1\over p \cdot q }\
{\red{ F_3(x_{\rm bj},Q^2)}}.
\label{DIS4}
\end{eqnarray}

If we combine Eqs.~(\ref{DIS1},\ref{DIS2},\ref{DIS3},\ref{DIS4}), we
can write the cross section for deeply inelastic scattering in terms
of the three structure functions.  Neglecting the hadron mass compared
to $Q^2$, the result is
\begin{equation}
{d \sigma \over dx_{\rm bj}\, dy} =
{\green{\tilde N(Q^2)}}\left[
y {\red{F_1}}
+ {1-y\over x_{\rm bj} y}
{\red{F_2}}
+ {\green{\delta_V}}\,(1-{y\over 2}) {\red{F_3}}
\right].
\label{DIS5}
\end{equation}
Here the normalization factor ${\green{\tilde N}}$ and the factor
$\delta_V$ multiplying $F_3$ are
\begin{eqnarray}
{\green{\tilde N}} &=& {4\pi \alpha^2 \over Q^2},
\hskip 3.42 cm
{\green{\delta_V}} = 0,
\hskip 0.55 cm
e^-\! + h \to e^- + X,
\nonumber\\
{\green{\tilde N}} &=& 
{\pi \alpha^2 Q^2\over 4 \sin^4\!(\theta_W)\  (Q^2 + M_W)^2},
\hskip 0.5 cm
{\green{\delta_V}} = 1,
\hskip 0.55 cm
\nu + h \to \mu^- + X,
\nonumber\\
{\green{\tilde N}} &=& {\pi \alpha^2 Q^2\over 4 \sin^4\!(\theta_W)\ 
(Q^2 + M_W)^2},
\hskip 0.5 cm
{\green{\delta_V}} = -1,
\hskip 0.2 cm
\bar\nu + h \to \mu^+ +X.
\end{eqnarray}
In principle, one can use the $y$ dependence to determine all three
of $F_1,F_2,F_3$ in a deeply inelastic scattering experiment.

\subsection{Space-time structure of DIS}

So far, we have used the symmetries of QCD in order to write the
cross section for deeply inelastic scattering in terms of three
structure functions, but we have not used any other dynamical
properties of the theory. Now we turn to the question of how the
scattering develops in space and time. 

\begin{figure}[htb]
\centerline{\DESepsf(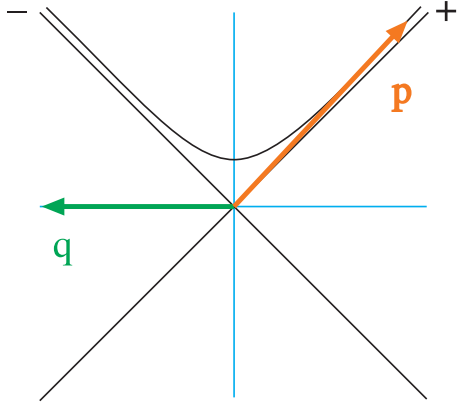 width 6 cm)}
\caption{Reference frame for the analysis of deeply inelastic
scattering.}
\label{DISframe}
\end{figure}

For this purpose, we define a convenient reference frame, which is
illustrated in Fig.~\ref{DISframe}. Denoting components of vectors
$v^\mu$ by
$(v^+,v^-,{\bf v}_T)$, we chose the frame in which
\begin{equation}
(q^+,q^-,{\bf q}) = {1 \over \sqrt 2}\
(-Q,Q,{\bf 0}).
\end{equation}
We also demand that the transverse components of the hadron momentum
be zero in our frame. Then
\begin{equation}
(p^+,p^-,{\bf p}) \approx {1 \over \sqrt 2}\
({Q \over x_{\rm bj}},{x_{\rm bj} m_h^2 \over Q},{\bf 0}).
\end{equation}
Notice that in the chosen reference frame the hadron momentum is
big and the momentum transfer is big.

\begin{figure}[htb]
\centerline{\DESepsf(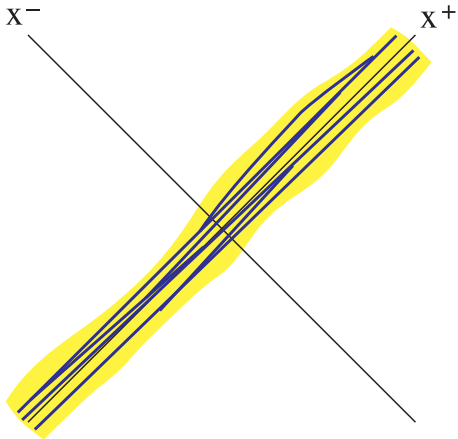 width 6 cm)}
\caption{Interactions within a fast moving hadron. The lines
represent world lines of quarks and gluons. The interaction points
are spread out in $x^+$ and pushed together in $x^-$.}
\label{DISB}
\end{figure}

Consider the interactions among the quarks and gluons inside a
hadron, using $x^+$ in the role of ``time'' as in Section
\ref{NullPlane}. For a hadron at rest, these interactions happen in a
typical time scale $\Delta x^+ \sim 1/m$, where $m \sim 300\ \MeV$.
A hadron that will participate in a deeply inelastic scattering event
has a large momentum, $p^+ \sim Q$, in the reference
frame that we are using.  The Lorentz transformation from the rest
frame spreads out interactions by a factor $Q/m$, so that
\begin{equation}
\Delta x^+ \sim {1 \over m}\times {\blue{{Q \over m}}} 
= {Q \over m^2}.
\end{equation}
This is illustrated in Fig.~\ref{DISB}.

I offer two caveats here. First, I am treating $x_{\rm bj}$ as being
of order 1. To treat small $x_{\rm bj}$ physics, one needs to put
back the factors of $x_{\rm bj}$, and the picture changes rather
dramatically. Second, the interactions among the quarks and gluons
in a hadron at rest can take place on time scales $\Delta x^+$ that
are much smaller than $1/m$, as we discussed in Section
\ref{smallest}. We will discuss this later on, but for now we start
with the simplest picture.

\begin{figure}[htb]
\centerline{\DESepsf(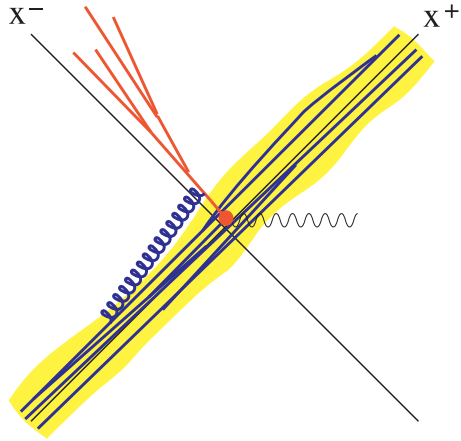 width 6 cm)}
\caption{The virtual photon meets the fast moving hadron. One of the
partons is annihilated and recreated as a parton with a large
minus component of momentum. This parton develops into a jet of
particles.}
\label{DISC}
\end{figure}

What happens when the fast moving hadron meets the virtual photon?
The interaction with the photon carrying momentum $q^- \sim Q$ is
localized to within
\begin{equation}
\Delta x^+ \sim {1 / Q}.
\end{equation}
During this short time interval, the quarks and gluons in the proton
are effectively free, since their typical interaction times are
comparatively much longer.

We thus have the following picture. At the moment $x^+$ of the
interaction, the hadron effectively consists of a collection of
quarks and gluons ({\it partons}) that have momenta $(p_i^+,{\bf
p}_i)$. We can treat the partons as being free. The $p_i^+$ are
large, and it is convenient to describe them using momentum fractions
$\xi_i$:
\begin{equation}
{\blue{\xi_i = p_i^+/p^+}},\hskip 1 cm 0<\xi_i < 1.
\end{equation}
(This is convenient because the $\xi_i$ are invariant under boosts
along the $z$ axis.) The transverse momenta of the partons, ${\bf
p}_i$, are small compared to $Q$ and can be neglected in the
kinematics of the $\gamma$-parton interaction. The ``on-shell'' or
``kinetic'' minus momenta of the partons, $p_i^- = {\bf p}_i^2/(2p_i
^+)$, are also very small compared to $Q$ and can be neglected in the
kinematics of the $\gamma$-parton interaction. We can think of the
partonic state as being described by a wave function
\begin{equation}
\psi(p_1^+, {\bf p}_1;p_2^+, {\bf p}_2;\cdots), 
\end{equation}
where indices specifying spin and flavor quantum numbers have been
suppressed.

\begin{figure}[htb]
\centerline{\DESepsf(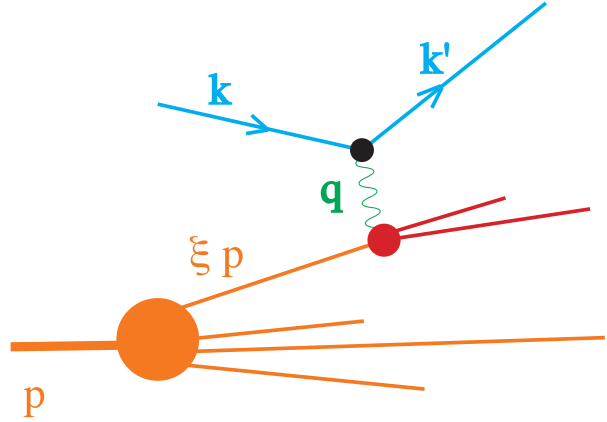 width 8 cm)}
\caption{Feynman diagram for deeply inelastic scattering.}
\label{DISD}
\end{figure}

This approximate picture is represented in Feynman diagram language
in Fig.~\ref{DISD}. The larger filled circle represents the hadron
wave function $\psi$. The smaller filled circle represents a sum of
subdiagrams in which the particles have virtualities of order $Q^2$.
All of these interactions are effectively instantaneous on the time
scale of the intra-hadron interactions that form the wave function.
The approximate picture also leads to an intuitive formula that
relates the observed cross section to the cross section for
$\gamma$-parton scattering: 
\begin{equation}
{{\green{d \sigma}} \over dE^\prime\, d\omega^\prime} \sim
\int_0^1\! d \xi \sum_a\ {\blue{f_{\! a/h}\!(\xi,\mu)}}\ 
{{\red{d \hat\sigma}}_{\!a}(\mu) \over dE^\prime\, d\omega^\prime}
+{\cal O}(m/Q).
\label{factor}
\end{equation}

In Eq.~(\ref{factor}), the function $f$ is a parton distribution
function: ${\blue{f_{\! a/h}\!(\xi,\mu)}}\  d\xi$ gives
probability to find a parton with flavor $a = g,u,\bar u, d, \dots$,
in hadron $h$, carrying momentum fraction within $d\xi$ of $\xi =
p_i^+/p^+$. If we knew the wave functions $\psi$, we would form $f$
by summing over the number $n$ of unobserved partons, integrating
$|\psi_n|^2$ over the momenta of the unobserved partons, and also
integrating over the transverse momentum of the observed parton. 

The second factor in Eq.~(\ref{factor}), ${{\red{d\hat \sigma}}_{\!a}
/ dE^\prime\, d\omega^\prime}$, is the cross section for scattering
the lepton from the parton of flavor $a$ and momentum fraction $\xi$.

I have indicated a dependence on a factorization scale $\mu$ in both
factors of Eq.~(\ref{factor}). This dependence arises from
the existence of virtual processes among the partons that take place
on a time scale much shorter than the nominal $\Delta x^+ \sim
Q/m^2$. I will discuss this dependence in some detail shortly.

\subsection{The hard scattering cross section}

\begin{figure}[htb]
\centerline{\DESepsf(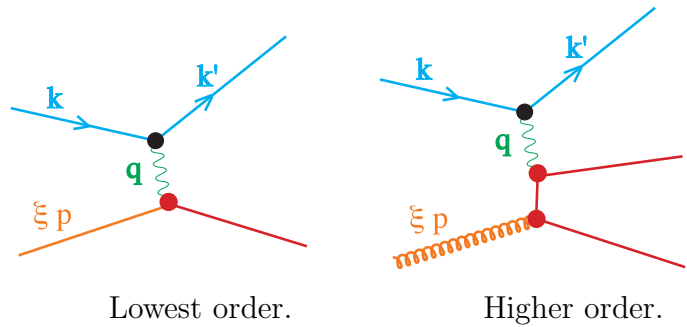 width 9 cm)}
\centerline{\hskip 0.4 in Lowest order.\hskip 1 in Higher order.}
\caption{Some Feynman diagrams for the hard scattering part of
deeply inelastic scattering.}
\label{DISE}
\end{figure}

The parton distribution functions in Eq.~(\ref{factor}) are derived
from experiment. The hard scattering cross sections ${{\red{d
\hat\sigma}}_{\!a}(\mu) / dE^\prime\, d\omega^\prime}$ are
calculated in perturbation theory, using diagrams like those shown
in Fig.~\ref{DISE}. The diagram on the left is the lowest order
diagram. The diagram on the right is one of several that contributes
to $d \hat\sigma$ at order $\alpha_s$; in this diagram the parton
$a$ is a gluon.

\begin{figure}[htb]
\centerline{\DESepsf(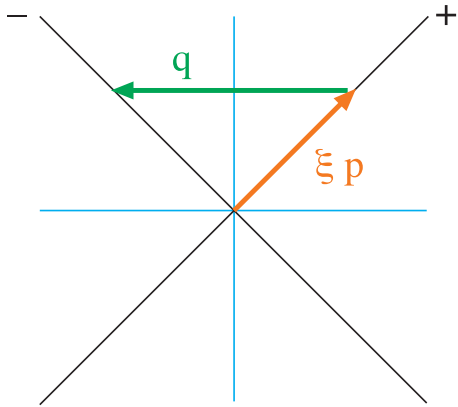 width 6 cm)}
\caption{Kinematics of lowest order diagram.}
\label{DISframeB}
\end{figure}

One can understand a lot about deeply inelastic scattering from
Fig.~\ref{DISframeB}, which illustrates the kinematics of the lowest
order diagram. Recall that in the reference frame that we are using,
the virtual vector boson has zero transverse momentum. The incoming
parton has momentum along the plus axis. After the scattering, the
parton momentum must be on the cone $k_\mu k^\mu = 0$, so the only
possibility is that its minus momentum is non-zero and its plus
momentum vanishes. That is 
\begin{equation}
\xi p^+ + q^+ = 0.
\end{equation}
Since $p^+ = Q/(x_{\rm bj}\sqrt 2)$ while $q^+ = - Q/\sqrt 2$, this
implies
\begin{equation}
{\blue{\xi = x_{\rm bj}}}.
\end{equation}
The consequence of this is that the lowest order contribution
to $d\hat \sigma$ in Eq.~(\ref{factor}) contains a delta function that
sets $\xi$ to $x_{\rm bj}$. Thus deeply inelastic scattering at a
given value of $x_{\rm bj}$ provides a determination of the
parton distribution functions at momentum fraction $\xi$ equal to
$x_{\rm bj}$, as long as one works only to leading order. In fact,
because of this close relationship, there is some tendency to confuse
the structure functions $F_n(x_{\rm bj}, Q^2)$ with the parton
distribution functions $f_{a,h}(\xi,\mu)$. I will try to keep these
concepts separate: the structure functions $F_n$ are something that
one measures directly in deeply inelastic scattering; the parton
distribution functions are determined rather indirectly from
experiments like deeply inelastic scattering, using formulas that are
correct only up to some finite order in
$\alpha_s$.

\subsection{Factorization for the structure functions}

We will look at DIS in a little detail since it is so important. Our
object is to derive a formula relating the measured structure functions
to structure functions calculated at the parton level. Then we will
look at the parton level calculation at lowest order. 

Start with Eq.~(\ref{factor}), representing Fig.~\ref{DISD}. We
change variables in this equation from $(E^\prime,\omega^\prime)$ to
$(x_{\rm bj},y)$. We relate $x_{\rm bj}$ to the momentum fraction
$\xi$ and a new variable $\hat x$ that is just $x_{\rm bj}$ with the
proton momentum $p^\mu$ replaced by the parton momentum $\xi p^\mu$: 
\begin{equation}
{\green{x_{\rm bj}}} = { Q^2 \over 2 p \cdot q}
= \xi\ { Q^2 \over 2 \xi p \cdot q}
={\blue{\xi \hat x}}.
\end{equation}
That is, $\hat x$ is the parton level version of $x_{\rm bj}$.
The variable $y$ is identical to the parton level version of $y$
because $p^\mu$ appears in both the numerator and denominator:
\begin{equation}
y = {p\cdot q \over p\cdot k}={\xi p\cdot q \over \xi p\cdot k} .
\end{equation}
Thus Eq.~(\ref{factor}) becomes
\begin{equation}
{d \sigma \over {\green{dx_{\rm bj}}}\, dy} \sim
\int_0^1\! d \xi \sum_a\ f_{\! a/h}\!(\xi)\ 
{1 \over {\blue{\xi}}}
\left[{d \hat\sigma_{\!a}\over
{\blue{d\hat x}}\,dy}\right]_{\hat x = x_{\rm bj}/\xi}
+{\cal O}(m/Q).
\label{factorA}
\end{equation}

Now recall that, for $\gamma$ exchange, $d\sigma/( dx_{\rm bj} dy)$ is
related to the structure functions by Eq.~(\ref{DIS5}):
\begin{equation}
{\green{{d \sigma\over dx_{\rm bj}\,dy}}}
=
\tilde N(Q^2)\!\left[
y\, {\green{F_1}}(x_{\rm bj},Q^2)
+ {1-y\over x_{\rm bj}\, y}\,
{\green{F_2}}(x_{\rm bj},Q^2)
\right]
+{\cal O}(m/Q).
\label{factorB}
\end{equation}
We define structure functions $\hat F_n$ for partons in the same way:
\begin{equation}
{\blue{{d \hat\sigma_{\!a}\over d\hat x\,dy}}}
=
\tilde N(Q^2)\left[
y\, {\blue{\hat F_1^a}}(x_{\rm bj}/\xi,Q^2)
+ {1-y\over (x_{\rm bj}/{\red{\xi}})y}\,
{\blue{\hat F_2^a}}(x_{\rm bj}/\xi,Q^2)
\right].
\label{factorC}
\end{equation}
We insert Eq.~(\ref{factorC}) into Eq.~(\ref{factorA}) and compare
to Eq.~(\ref{factorB}).  We deduce that the structure functions can be
factored as
\begin{equation}
{\green{F_1}}(x_{\rm bj},Q^2) \sim
\int_0^1\! {\red{d \xi}} \sum_a\ f_{\! a/h}\!(\xi)\ 
{1 \over {\red{\xi}}}
{\blue{\hat F_1^a}}(x_{\rm bj}/\xi,Q^2)
+{\cal O}(m/Q),
\label{factorF1}
\end{equation}
\begin{equation}
{\green{F_2}}(x_{\rm bj},Q^2) \sim
\int_0^1\! {\red{d \xi}} \sum_a\ f_{\! a/h}\!(\xi)\ 
\hat {\blue{F_2^a}}(x_{\rm bj}/\xi,Q^2)
+{\cal O}(m/Q).
\label{factorF2}
\end{equation}

A simple calculation gives $\hat F_1$ and $\hat F_2$ at lowest order:
\begin{equation}
{\blue{\hat F_1^a}}(x_{\rm bj}/\xi,Q^2) =
{1\over 2}Q_a^2\ \delta(x_{\rm bj}/\xi - 1) + {\cal O}(\alpha_s),
\end{equation}
\begin{equation}
{\blue{\hat F_2^a}}(x_{\rm bj}/\xi,Q^2) =
Q_a^2\ \delta(x_{\rm bj}/\xi - 1) + {\cal O}(\alpha_s).
\end{equation}
Inserting these results into Eqs.~(\ref{factorF1}) and
(\ref{factorF2}), we obtain the lowest order relation between the
structure functions and the parton distribution functions:
\begin{equation}
{\green{F_1}}(x_{\rm bj},Q^2) \sim
{1\over 2}\sum_a\ Q_a^2\ {\red{f_{\! a/h}\!(x_{\rm bj})}}
+ {\cal O}(\alpha_s) +{\cal O}(m/Q),
\end{equation}
\begin{equation}
{\green{F_2}}(x_{\rm bj},Q^2) \sim
\sum_a\ Q_a^2\ {\red{x_{\rm bj}\,f_{\! a/h}\!(x_{\rm bj})}}
+ {\cal O}(\alpha_s) +{\cal O}(m/Q).
\end{equation}
The factor 1/2 between $x_{\rm bj} F_1$ and $F_2$ follows from the
Feynman diagrams for spin 1/2 quarks. 

\subsection{$\mu_{F}$ dependence}

\begin{figure}[htb]
\centerline{\DESepsf(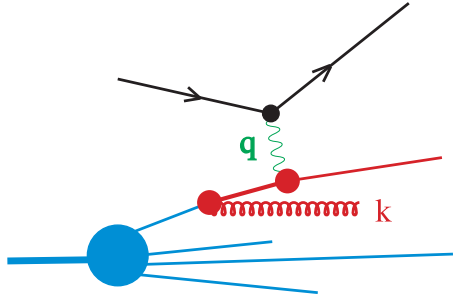 width 6 cm)}
\caption{Deeply inelastic scattering with a gluon emission.}
\label{DISF}
\end{figure}

I have so far presented a rather simplified picture of deeply
inelastic scattering in which the hard scattering takes place on a
time scale $\Delta x^+ \sim 1/Q$, while the internal dynamics of the
proton take place on a much longer time scale $\Delta x^+ \sim
Q/m^2$. What happens when one actually computes Feynman diagrams and
looks at what time scales contribute?  Consider the graph shown in
Fig.~\ref{DISF}. One finds that the transverse momenta ${\bf k}$
range from order $m$ to order $Q$, corresponding to  energy scales
$k^- = {\bf k}^2/2k^+$ between $k^- \sim m^2/Q$ and $k^- = Q^2/Q \sim
Q$, or time scales $Q/m^2 \lesssim \Delta x^+ \lesssim 1/Q$. 

The property of factorization for the cross section of deeply
inelastic scattering, embodied in Eq.~(\ref{factor}), is established
by showing that the perturbative expansion can be rearranged so that
the contributions from long time scales appear in the parton
distribution functions, while the contributions from short time
scales appear in the hard scattering functions. (See
Ref.~\citenum{factor} for more information.) Thus, in 
Fig.~\ref{DISF}, a gluon emission with ${\bf k}^2 \sim m^2$ is part of
${\blue{f(\xi)}}$, while a gluon emission with ${\bf k}^2 \sim Q^2$ is
part of ${\red{d \hat\sigma}}$.

\begin{figure}[htb]
\centerline{\DESepsf(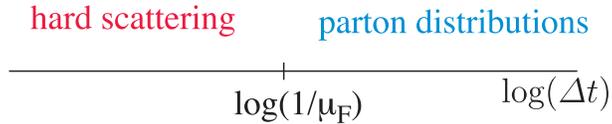 width 8 cm)}
\caption{Distance scales in factorization.}
\label{scalesB}
\end{figure}

Breaking up the cross section into factors associated with short and
long time scales requires the introduction of a {\it factorization
scale}, $\mu_F$. When calculating the diagram in Fig.~\ref{DISF}, one
integrates over $\bf k$. Roughly speaking, one counts the
contribution from ${\bf k}^2 < \mu_{F}^2$ as part of  the higher
order contribution to $\phi(\xi)$, convoluted with the lowest order
hard scattering function $d\hat\sigma$ for deeply inelastic
scattering from a quark. The contribution from  $\mu_{F}^2 < {\bf
k}^2$ then counts as part of the higher order contribution to
$d\hat\sigma$ convoluted with an uncorrected parton distribution.
This is illustrated in Fig.~\ref{scalesB}. (In real calculations, the
split is accomplished with the aid of dimensional regularization, and
is a little more subtle than a simple division of the integral into
two parts.)

A consequence of this is that both ${\red{{d
\hat\sigma_{\!a}(\mu_{\!F}) / dE^\prime\,  d\omega^\prime}}}$\ and 
${\blue{f_{\!a/h}(\xi,\mu_{\!F})}}$  depend on $\mu_{\!F}$.
Thus we have two scales, the factorization scale $\mu_{F}$ in
$f_{\!f/h}(\xi,\mu_{\!F})$ and the renormalization scale $\mu$ in
$\alpha_s(\mu)$. As with $\mu$, the cross section does not depend on
$\mu_F$. Thus there is an equation $d ({\it cross\ section})/d\mu_F =
0$ that is satisfied to the accuracy of the perturbative calculation
used. If you work harder and calculate to higher order, then the
dependence on $\mu_F$ is less.

Often one sets $\mu_F = \mu$ in applied calculations. In fact, it is
rather common in applications to deeply inelastic scattering to set 
$\mu_F = \mu = Q$.

\subsection{Contour graphs of scale dependence}

As an example, look at the one jet inclusive cross section in
proton-antiproton collisions. Specifically, consider the cross section
$d\sigma/d E_T d\eta$ to make a collimated spray of particles, a {\it
jet}, with transverse energy $E_T$  and rapidity $\eta$. (Here $E_T
$ is essentially the transverse momentum carried by the particles in
the jet and $\eta$ is related to the angle between the jet and
the beam direction by $\eta \equiv \ln(\tan(\theta/2)$).  We will
investigate this process and discuss the definitions in the next
section. For now, all we need to know is that the theoretical formula
for the cross section at next-to-leading order involves the strong
coupling $\alpha_s(\mu)$ and two factors $f_{a/h}(x,\mu_F)$
representing the distribution of partons in the two incoming hadrons.
There is a parton level hard scattering cross section that also
depends on $\mu$ and $\mu_F$.

\begin{figure}[htb]
\centerline{\DESepsf(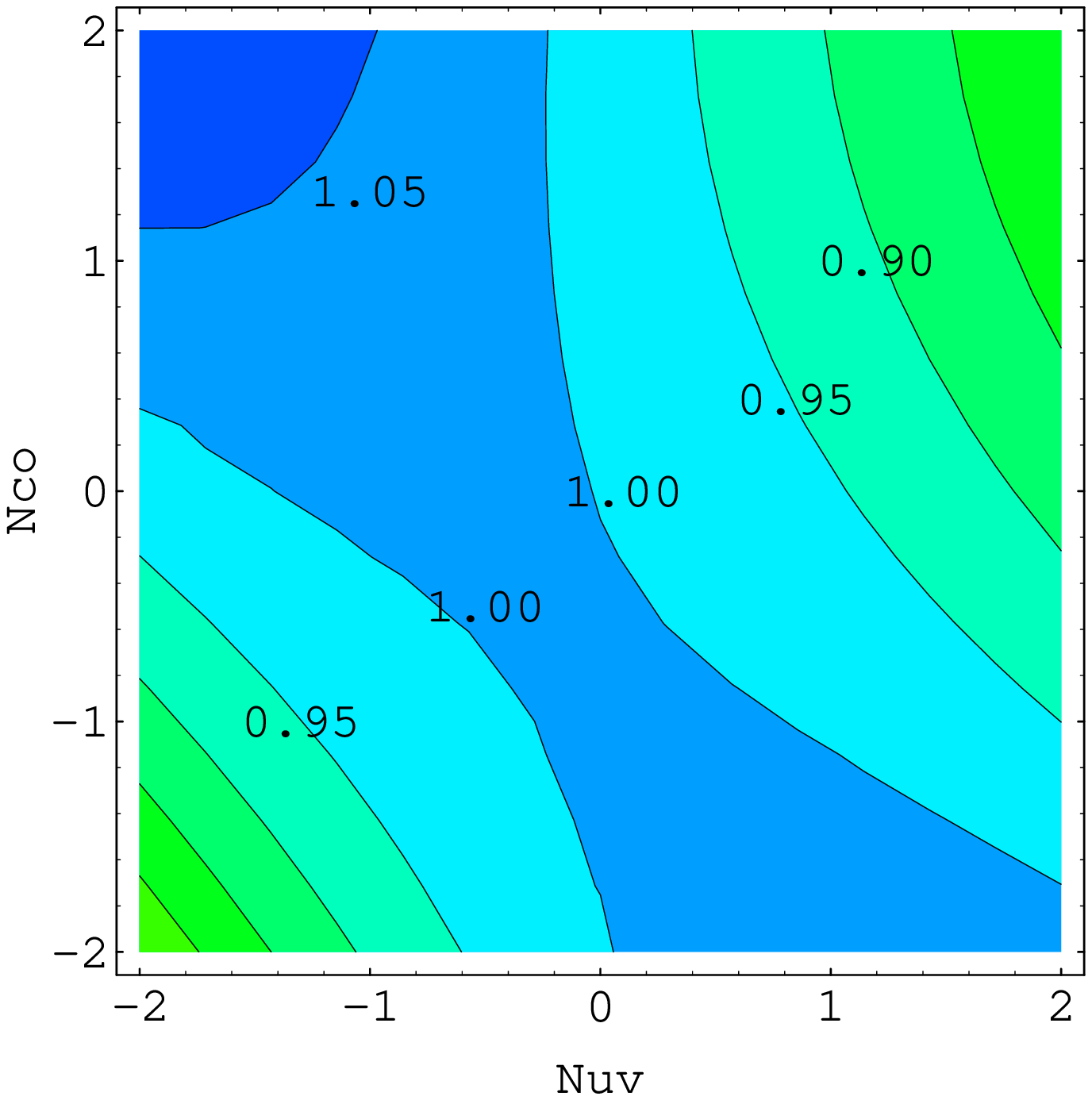 width 7 cm)
            \DESepsf(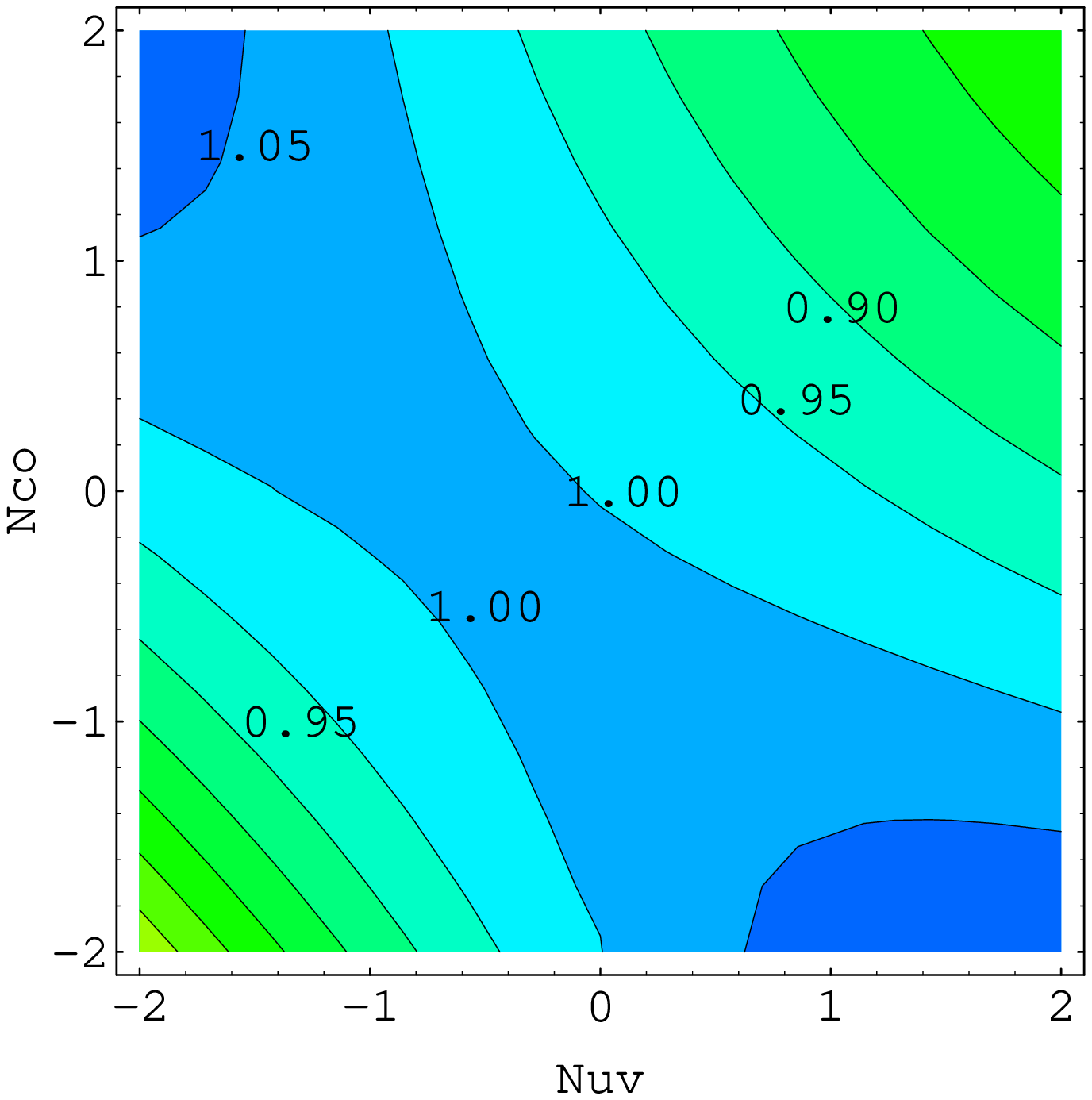 width 7 cm)}
\centerline{\hskip 1 cm
$E_T = 100\GeV$\hskip 4 cm
$E_T = 500\GeV$}
\caption{Contour plots of the one jet inclusive cross section
versus the renormalization scale
$\mu$ and the factorization scale $\mu_F$. The cross section is
$d\sigma / dE_T d\eta$ at $\eta = 0$ with $E_T = 100\ \GeV$ in
the first graph and  $E_T = 500\ \GeV$ in the second.  The horizontal
axis in each graph represents $N_{UV} \equiv \log_2(2\mu/E_T)$ and the
vertical axis represents  $N_{CO} \equiv \log_2(2\mu_F/E_T)$. The
contour lines show 5\% changes in the cross section relative to the
cross section at the center of the figures. The c.m energy is
$\protect\sqrt s = 1800\
\GeV$.}
\label{mu100500}
\end{figure}

How does the cross section depend on $\mu$ in $\alpha_s(\mu)$ and
$\mu_F$ in $f_{a/h}(x,\mu_F)$? In Fig.~\ref{mu100500}, I show contour
plots of the jet cross section versus $\mu$ and $\mu_F$ at two
different values of $E_T$. The center of the plots corresponds to a
standard choice of scales, $\mu = \mu_F = E_T/2$. The axes are
logarithmic, representing $\log_2(2\mu/E_T)$ and $\log_2(2\mu_F/E_T)$.
Thus $\mu$ and $\mu_F$ vary from $E_T/8$ to $2 E_T$ in the plots.

Notice that the dependence on the two scales is rather mild for the
next-to-leading order cross section. The cross section
calculated at leading order is quite sensitive to these scales,
but most of the scale dependence found at order $\alpha_s^2$ has been
canceled by the $\alpha_s^3$ contributions to the cross section.
One reads from the figure that the cross section varies by
roughly ${\blue{\pm 15\%}}$ in the central region of the graphs, both
for medium and large $E_T$. Following the argument of
Sec.~\ref{errorest}, this leads to a rough estimate of 15\% for the
theoretical error associated with truncating perturbation theory at
next-to-leading order.

\subsection{\MSbar\ definition of parton distribution functions}

The factorization property, Eq.~(\ref{factor}), of the deeply
inelastic scattering cross section states that the cross section
can be approximated as a convolution of a hard scattering cross
section that can be calculated perturbatively and parton
distribution functions $f_{a/A}(x,\mu)$. But what are the parton
distribution functions? This question has some practical importance.
The hard scattering cross section is essentially the physical
cross section divided by the parton distribution function, so the
precise definition of the parton distribution functions leads to the
rules for calculating the hard scattering functions.

The definition of the parton distribution functions is to some
extent a matter of convention. The most commonly used convention is
the \MSbar\ definition, which arose from the theory of deeply
inelastic scattering in the language of the ``operator product
expansion.''\cite{msbar} Here I will follow the
(equivalent) formulation of Ref.~\citenum{CSparton}. For a more
detailed pedagogical review, the reader may consult
Ref.~\citenum{DESlattice}.

Using the \MSbar\ definition, the distribution of quarks in a hadron
is given as the hadron matrix element of certain quark field
operators:
\begin{equation}
{\blue{f_{i/h}(\xi,\mu_F)}}=
{1 \over 2}\int {dy^- \over 2\pi}\ e^{-i \xi p^+ y^-}
\langle p| {\blue{\bar\psi}}_i(0,y^-,{\bf 0}) \gamma ^+ 
{\magenta{F}}
{\blue{\psi}}_i(0)|p\rangle.
\end{equation}
Here $|p\rangle$ represents the state of a hadron with
momentum $p^\mu$ aligned so that  $p_T = 0$. For simplicity,
I take the hadron to have spin zero.  The operator $\psi(0)$,
evaluated at $x^\mu = 0$, annihilates a quark in the hadron. The
operator ${\blue{\bar\psi}}_i(0,y^-,{\bf 0})$ recreates the quark at
$x^+ = {\bf x}_T = 0$ and $x^- = y^-$, where we take the appropriate
Fourier transform in $y^-$ so that the quark that was annihilated and
recreated has momentum $k^+ = \xi p^+$. The motivation for the
definition is that this is the hadron matrix element of the
appropriate number operator for finding a quark. 
 
There is one subtle point. The number operator idea corresponds to a
particular gauge choice, $A^+ = 0$. If we are using any other gauge,
we insert the operator
\begin{equation}
{\magenta{F}}=
{\cal P}\exp\left(
-ig\int_0^{y^-} dz^- {\blue{A}}_a^+(0,z^-,{\bf 0})\, t_a
\right).
\end{equation}
The ${\cal P}$ indicates a path ordering of the operators and
color matrices along the path from $(0,0,{\bf 0})$ to $(0,y^-,{\bf
0})$. This operator is the identity operator in $A^+ = 0$ gauge and
it makes the definition gauge invariant.

\begin{figure}[htb]
\centerline{\DESepsf(DISC.eps width 6 cm)
\DESepsf(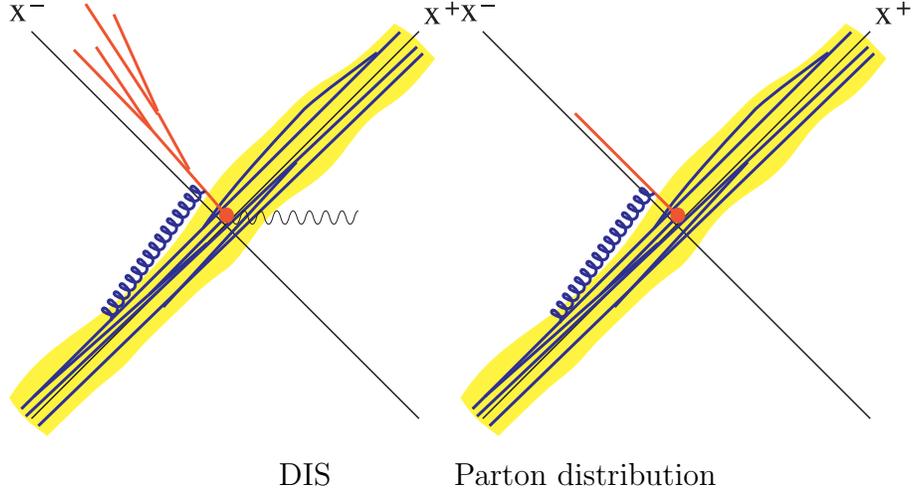 width 6 cm)}
\centerline{\hskip 1 cm DIS \hskip 1.5 cm Parton distribution}
\caption{Deeply inelastic scattering and the parton distribution
functions.}
\label{DISG}
\end{figure}

The physics of this definition is illustrated in Fig.~\ref{DISG}. The
first picture (from Fig.~\ref{DISC}) illustrates the amplitude for
deeply inelastic scattering. The fast proton moves in the plus
direction. A virtual photon knocks out a quark, which emerges moving
in the minus direction and develops into a jet of particles. The
second picture illustrates the amplitude associated with the quark
distribution function. We express $F$ as $F_2 F_1$ where
\begin{eqnarray}
{\magenta{F_2}} &=&
\bar{\cal P}\exp\left(
+ig\int^\infty_{y^-}\!\! dz^- {\blue{A}}_a^+(0,z^-,{\bf 0})\, t_a
\right),
\nonumber\\
{\magenta{F_1}} &=&
{\cal P}\exp\left(
-ig\int_0^{\infty}\!\! dz^- {\blue{A}}_a^+(0,z^-,{\bf 0})\, t_a
\right).
\end{eqnarray}
and write the quark distribution function including a sum over
intermediate states $|N\rangle$:
\begin{equation}
{\blue{f_{i/h}(\xi,\mu_F)}}=
{1 \over 2}\int {dy^- \over 2\pi}\ e^{-i \xi p^+ y^-}
\sum_N
\langle p| {\blue{\bar\psi}}_i(0,y^-,{\bf 0}) \gamma ^+ 
{\magenta{F_2}}
|N\rangle \langle N|
{\magenta{F_1}}
{\blue{\psi}}_i(0)|p\rangle.
\end{equation}
Then the amplitude depicted in the second picture in Fig.~\ref{DISG}
is $\langle N|{\magenta{F_1}} {\blue{\psi}}_i(0)|p\rangle$. The
operator $\psi$ annihilates a quark in the proton. The operator
$F_1$ stands in for the quark moving in the minus direction. The gluon
field $A$ evaluated along a lightlike line in the minus direction
absorbs longitudinally polarized gluons from the color field of the
proton, just as the real quark in deeply inelastic scattering can
do. Thus the physics of deeply inelastic scattering is built into
the definition of the quark distribution function, albeit in an
idealized way. The idealization is not a problem because the hard
scattering function $d \hat \sigma$ systematically corrects for the
difference between real deeply inelastic scattering and the
idealization.

There is one small hitch. If you calculate any Feynman diagrams for 
${\blue{f_{i/h}(\xi,\mu_F)}}$, you are likely to wind up with an
ultraviolet-divergent integral. The operator product that is part of
the definition needs renormalization. This hitch is only a small
one. We simply agree to do all of the renormalization using the
\MSbar\ scheme for renormalization. It is this renormalization that
introduces the scale $\mu_F$ into ${\blue{f_{i/h}(\xi,\mu_F)}}$.
This role of $\mu_F$ is in accord with Fig.~\ref{scalesB}: roughly
speaking $\mu_F$ is the upper cutoff for what momenta belong with
the parton distribution function; at the same time it is the lower
cutoff for what momenta belong with the hard scattering function.

What about gluons? The definition of the gluon distribution function
is similar to the definition for quarks. We simply replace the quark
field $\psi$ by suitable combinations of the gluon field $A^\mu$, as
described in Refs.~\citenum{CSparton} and \citenum{DESlattice}.

\subsection{Evolution of the parton distributions}

Since we introduced a scale $\mu_F$ in the definition of the
parton distributions in order to define their renormalization,
there is a renormalization group equation that gives the $\mu_F$
dependence 
\begin{equation}
{d \over d \ln \mu_F}f_{a/h}(x,\mu_F) =
 \sum_b \int_x^1
{d \xi \over \xi}\
P_{ab}(x/\xi,\alpha_s(\mu_F))\
f_{b/h}(\xi,\mu_F).
\label{AP}
\end{equation}
This is variously known as the evolution equation, the
Altarelli-Parisi equation, and the DGLAP
(Dokshitzer-Gribov-Lipatov-Altarelli-Parisi) equation. Note the sum
over parton flavor indices. The  evolution of, say, an up quark ($a =
u$) can involve a gluon ($b = g$) through the element $P_{ug}$ of the
kernel that describes gluon splitting into $\bar u u$.

\begin{figure}[htb]
\centerline{\DESepsf(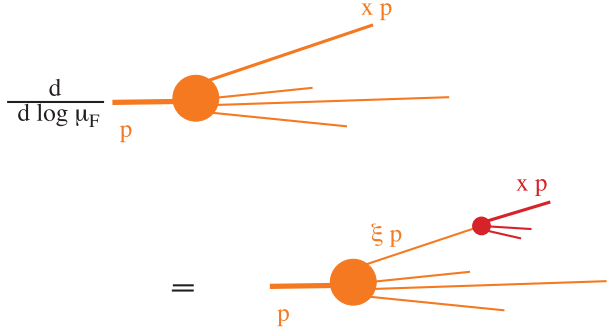 width 8 cm)}
\caption{The renormalization for the parton distribution functions.}
\label{APeqn}
\end{figure}

The equation is illustrated in Fig.~\ref{APeqn}. When we change the
renormalization scale $\mu_F$, the change in the probability to find a
parton with momentum fraction $x$ and flavor $a$ is proportional to
the probability to find such a parton with large transverse momentum.
The way to get this parton with large transverse momentum is for a
parton carrying momentum fraction $\xi$ and much smaller transverse
momentum  to split into partons carrying large transverse momenta,
including the parton that we are looking for. This splitting
probability, integrated over the appropriate transverse momentum
ranges, is the kernel $P_{ab}$.

The kernel $P$ in Eq.~(\ref{AP}) has a perturbative expansion
\begin{equation}
P_{ab}(x/\xi,\alpha_s(\mu_F)) =
P_{ab}^{(1)}(x/\xi)\ 
{\alpha_s(\mu_F) \over \pi}
+
P_{ab}^{(2)}(x/\xi)\ 
\left({\alpha_s(\mu_F) \over \pi}\right)^2
+\cdots.
\end{equation}
The first two terms are known and are typically used in numerical
solutions of the equation. To learn more about the DGLAP equation,
the reader may consult Refs.~\citenum{handbook} and
\citenum{DESlattice}.

\subsection{Determination and use of the parton distributions}

The \MSbar\ definition giving the parton distribution in terms of
operators is process independent -- it does not refer to any
particular physical process. These parton distributions then appear in
the QCD formula for {\blue{any process with one or two hadrons in the
initial state}}. In principle, the parton distribution functions
could be calculated by using the method of lattice QCD (see
Ref.~\citenum{DESlattice}). Currently, they are determined from
experiment.

Currently the most comprehensive analyses are being done by the CTEQ
\cite{CTEQ4} and MRS \cite{MRSpartons} groups. These groups
perform a ``global fit'' to data from experiments of several
different types. To perform such a fit one chooses a parameterization
for the parton distributions at some standard factorization scale
$\mu_0$. Certain sum rules that follow from the definition of the
parton distribution functions are built into the parameterization. An
example is the momentum sum rule:
\begin{equation}
\sum_a \int_0^1 d\xi\ \xi\ f_{a/h}(\xi,\mu) = 1.
\end{equation}
Given some set of values for the parameters describing the
$f_{a/h}(x,\mu_0)$, one can determine $f_{a/h}(x,\mu)$ for all
higher values of $\mu$ by using the evolution equation. Then the QCD
cross section formulas give predictions for all of the experiments
that are being used. One systematically varies the parameters in
$f_{a/h}(x,\mu_0)$ to obtain the {\it best} fit to all of the
experiments. One source of information about these fits is the
World Wide Web pages of Ref.~\citenum{potpourri}.

If the freedom available for the parton distributions is used to fit
all of the world's data, is there any physical content to QCD? The
answer is yes:  there are lots of experiments, so this program won't
work unless QCD is right. In fact, there are roughly 1400 data in the
CTEQ fit and only about 25 parameters available to fit these data.

\section{ QCD in hadron-hadron collisions}
 
When there is a hadron in the initial state of a scattering process,
there are inevitably long time scales associated with the binding
of the hadron, even if part of the process is a short-time
scattering. We have seen, in the case of deeply inelastic
scattering of a lepton from a single hadron, that the dependence on
these long time scales can be factored into a parton distribution
function. But what happens when two high energy hadrons collide? The
reader will not be surprised to learn that we then need two parton
distribution functions. 

I explore hadron-hadron collisions in this section. I begin with the
definition of a convenient kinematical variable, rapidity. Then I
discuss, in turn, production of vector bosons ($\gamma^*$, $W$, and
$Z$), heavy quark production, and jet production.

\subsection{Kinematics: rapidity}

\begin{figure}[htb]
\centerline{\DESepsf(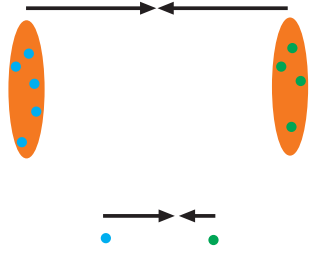 width 4 cm)}
\caption{Collision of two hadrons containing partons producing a $Z$
boson. The c.m.\ frame of the two hadrons is normally $\it not$ the
c.m.\ frame of the two partons that create the $Z$ boson.}
\label{ppcollision}
\end{figure}

In describing hadron-hadron collisions, it is useful to employ a
kinematic variable $y$ that is called {\it rapidity}.  Consider, for
example, the production of a $Z$ boson plus anything, $p + \bar p \to
Z + X$. Choose the hadron-hadron c.m.\ frame with the $z$ axis along
the beam direction. In Fig.~\ref{ppcollision}, I show a drawing of
the collision. The arrows represent the momenta of the two hadrons;
in the c.m.\ frame these momenta have equal magnitudes. We will want
to describe the process at the parton level, $a + b \to Z + X$.
The two partons $a$ and $b$ each carry some share of the parent
hadron's momentum, but generally these will not be equal shares.
Thus the magnitudes of the momenta of the colliding partons will not
be equal. We will have to boost along the $z$ axis in order to get
to the parton-parton c.m.\ frame. For this reason, it is useful to use
a variable that transforms simply under boosts. This is the
motivation for using rapidity.

Let $q^\mu = (q^+,q^-,{\bf q})$ be the momentum
of the $Z$ boson. Then the rapidity of the $Z$ is defined as
\begin{equation}
y = {1 \over 2} \ln\!\left({q^+ \over q^-}\right).
\end{equation}
The four components $(q^+,q^-,{\bf q})$ of the $Z$ boson momentum can
be written in terms of four variables, the two components of the $Z$
boson's transverse momentum ${\bf q}$, its mass $M$, and its rapidity:
\begin{equation}
q^\mu = (e^y\sqrt{({\bf q}^2 + M^2)/2},\   
e^{-y}\sqrt{({\bf q}^2 + M^2)/2}
,\ {\bf q}).
\end{equation}

The utility of using rapidity as one of the variables stems from the
transformation property of rapidity under a boost along the $z$ axis:
\begin{equation}
q^+ \to e^\omega q^+,\quad
q^- \to e^{-\omega} q^-,\quad
{\bf q} \to {\bf q}.
\end{equation}
Under this transformation, 
\begin{equation}
y \to y + \omega.
\end{equation}
This is as simple a transformation law as we could hope for. In
fact, it is just the same as the transformation law for velocities
in non-relativistic physics in one dimension.

\begin{figure}[htb]
\centerline{\DESepsf(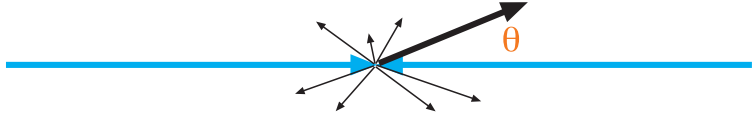 width 10 cm)}
\caption{Definition of the polar angle $\theta$ used in calculating
the rapidity of a massless particle.}
\label{rapidity}
\end{figure}

Consider now the rapidity of a {\blue{massless particle}}. Let the
massless particle emerge from the collision with polar angle
$\theta$, as indicated in Fig.~\ref{rapidity}. A simple calculation
relates the particle's rapidity $y$ to $\theta$:
\begin{equation}
y = -\ln\left(\tan(\theta/2)\right) \,,
\hskip 1 cm (m=0).
\end{equation}
Another way of writing this is
\begin{equation}
\tan \theta = 1/\sinh y \,,
\hskip 1 cm (m=0).
\end{equation}

One also defines the {\it pseudorapidity} $\eta$ of a particle,
massless or not, by
\begin{equation}
\eta = -\ln\left(\tan(\theta/2)\right) 
\hskip 1 cm {\rm or} \hskip 1 cm
\tan \theta = 1/\sinh \eta.
\end{equation}
The relation between rapidity and pseudorapidity is
\begin{equation}
\sinh \eta = \sqrt{1 + m^2/q_T^2}\ \sinh y.
\end{equation}
Thus, if the particle isn't quite massless,
$\eta$ may still be a good approximation
to $y$.

\subsection{$\gamma^*$, $W$, $Z$ production in hadron-hadron
collisions}
\
Consider the process 
\begin{equation}
A + B \to Z + X,
\label{makeaZ}
\end{equation}
where $A$ and $B$ are high energy hadrons. Two features of this
reaction are important for our discussion. First, the
mass of the $Z$ boson is large compared to 1 GeV, so that a process
with a small time scale $\Delta t \sim 1/M_Z$ must be involved in the
production of the $Z$. At lowest order in the strong interactions,
the process is $q + \bar q \to Z$. Here the quark and antiquark are
constituents of the high energy hadrons. The second significant
feature is that the $Z$ boson does not participate in the strong
interactions, so that our description of the observed final state can
be very simple.

We could equally well talk about $A + B \to W + X$ or $A + B \to
\gamma^* + X$ where the virtual photon decays into a muon pair or an
electron pair that is observed and where the mass of the $\gamma^*$ is
large compared to 1 GeV. This last process \cite{DrellYan}, $A + B \to
\gamma^* + X \to \ell^+ + \ell^- + X$, is historically important
because it helped establish the parton picture as being correct. The
$W$ and $Z$ processes were observed later. In fact, these are the
processes by which the $W$ and $Z$ bosons were first directly
observed \cite{WandZ}.

In process (\ref{makeaZ}), we allow the $Z$ boson to have any
transverse momentum ${\bf q}$. (Typically, then, ${\bf q}$ will be
much smaller than $M_Z$.) Since we integrate over $\bf q$ and the
mass of the $Z$ boson is fixed, there is only one variable needed to
describe the momentum of the $Z$ boson. We choose to use its
rapidity $y$, so that we are interested in the cross section $d
\sigma / dy$.

\begin{figure}[htb]
\centerline{\DESepsf(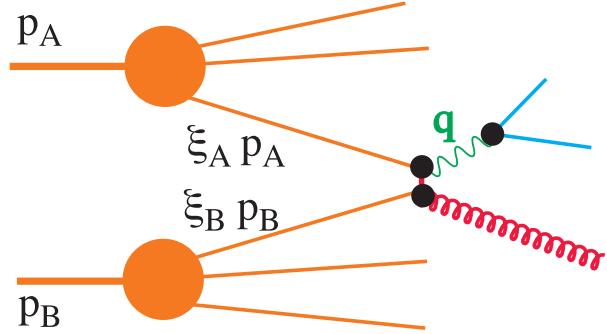 width 8 cm)}
\caption{A Feynman diagram for $Z$ boson production in a
hadron-hadron collision. Two partons, carrying momentum fractions
$\xi_A$ and $\xi_B$, participate in the hard interaction. This
particular Feynman diagram illustrates an order $\alpha_s$
contribution to the hard scattering cross section: a gluon
is emitted in the process of making the $Z$ boson. The diagram also
shows the decay of the $Z$ boson into an electron and a neutrino. }
\label{drellyan}
\end{figure}

The cross section takes a factored form similar to that found for
deeply inelastic scattering. Here, however, there are two parton
distribution functions:
\begin{equation}
{\green{{d \sigma \over d y}}}
\approx\! \sum_{a,b} \int_{x_A}^1\! d \xi_A \int_{x_B}^1\! d \xi_B\
{\blue{f_{a/A}(\xi_A,\mu_F)}}\ {\blue{f_{b/B}(\xi_B,\mu_F)}}\
{\red{{d \hat\sigma_{ab}(\mu,\mu_F) \over d y}}}.
\label{DYfactors}
\end{equation}
The meaning of this formula is intuitive:
${\blue{f_{a/A}(\xi_A,\mu_F)}}\,d\xi_A$ gives the probability to find
a parton in hadron $A$;
${\blue{f_{b/B}(\xi_B,\mu_f)}}\,d\xi_B$ gives the probability to find
a parton in hadron $B$;
${\red{{d \hat\sigma_{ab} / d y}}}$ gives the cross section for
these partons to produce the observed $Z$ boson. The formula is
illustrated in Fig.~\ref{drellyan}. The hard scattering cross section
can be calculated perturbatively. Fig.~\ref{drellyan} illustrates one
particular order $\alpha_s$ contribution to ${\red{{d \hat\sigma_{ab}
/ d y}}}$. The integrations over parton momentum fractions have limits
$x_A$ and $x_B$, which are given by
\begin{equation}
x_A = e^y \sqrt{M^2/s}, \quad\quad
x_B = e^{-y} \sqrt{M^2/s}.
\end{equation}

Eq.~(\ref{DYfactors}) has corrections of order $m/M_Z$, where $m$ is
a mass characteristic of hadronic systems, say 1 GeV. In addition,
when ${\red{{d \hat\sigma_{ab} / d y}}}$ is calculated to order
$\alpha_s^N$, then there are corrections of order $\alpha_s^{N+1}$.

There can be soft interactions between the partons in hadron
$A$ and the partons in hadron $B$, and these soft interactions can
occur before the hard interaction that creates the $Z$ boson. It
would seem that these soft interactions do not fit into the
intuitive picture that comes along with  Eq.~(\ref{DYfactors}). It is
a significant part of the factorization property that these soft
interactions do not modify the formula. These introductory lectures
are not the place to go into how this can be. For more information,
the reader is invited to consult Ref.~\citenum{factor}.

\subsection{Heavy quark production}

\begin{figure}[htb]
\centerline{\DESepsf(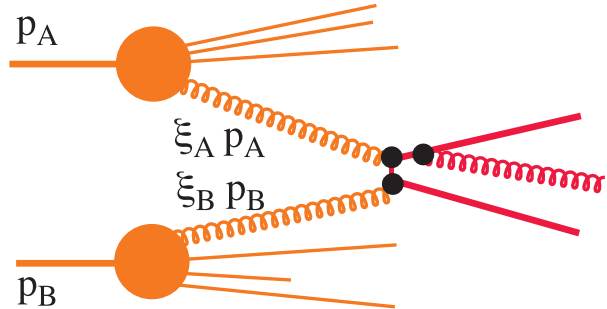 width 8 cm)}
\caption{Feynman graph for heavy quark production. The lowest order
hard process is $g + g \to Q + \bar Q$, which occurs at order
$\alpha_s^2$. This particular Feynman diagram illustrates an order
$\alpha_s^3$ process in which a gluon is emitted.}
\label{heavyquark}
\end{figure}

We now turn to the production of a heavy quark and its
corresponding antiquark in a high energy hadron-hadron collision:
\begin{equation}
A + B \to Q+ \bar Q + X.
\end{equation}
The most notable example of this is top quark production. A Feynman
diagram for this process is illustrated in Fig.~\ref{heavyquark}.

The total heavy quark production cross section takes a factored form
similar to that for $Z$ boson production, 
\begin{equation}
{\green{\sigma_T}} \approx\! 
\sum_{a,b} \int_{x_A}^1\! d \xi_A \int_{x_B}^1\! d \xi_B\
{\blue{f_{a/A}(\xi_A,\mu_F)}}\ {\blue{f_{b/B}(\xi_B,\mu_F)}}\
{\red{\hat\sigma_T}}^{ab}(\mu_F,\mu).
\label{heavyQfactors}
\end{equation}
As in the case of $Z$ production, $Q \bar Q$ production is a hard
process, with a time scale determined by the mass of the quark:
$\Delta t \sim 1/M_Q$. It is this hard process that is represented
by the calculated cross section ${\red{\hat\sigma_T}}^{ab}$. Of
course, the heavy quark and antiquark have strong interactions, and
can radiate soft gluons or exchange them with their environment.
These effects do not, however, affect the cross section: once the
$Q \bar Q$ pair is made, it is made. The probabilities for it to
interact in various ways must add to one. For an argument that 
Eq.~(\ref{heavyQfactors}) is correct, see Ref.~\citenum{CSSheavyQ}.

\subsection{Jet production}
\label{jetproduction}

In our study of high energy electron-positron annihilation, we
discovered three things. First, QCD makes the qualitative prediction
that particles in the final state should tend to be grouped in
collimated sprays of hadrons called jets. The jets carry the momenta
of the first quarks and gluons produced in the hard process.
Second, certain kinds of experimental measurements probe the
short-time physics of the hard interaction, while being insensitive to
the long-time physics of parton splitting, soft gluon exchange, and
the binding of partons into hadrons. Such measurements are called
infrared safe. Third, among the infrared safe observables are cross
sections to make jets.

\begin{figure}[htb]
\centerline{ \DESepsf(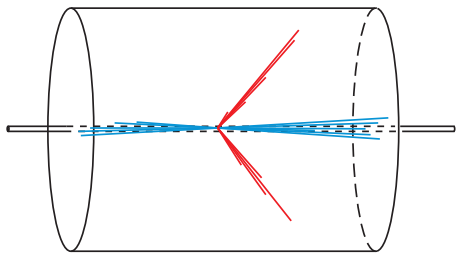 width 6 cm) }
\caption{Sketch of a two-jet event at a hadron collider. The
cylinder represents the detector, with the beam pipe along its
axis. Typical hadron-hadron collisions produce beam remnants, the
debris from soft interactions among the partons. The particles in
the beam remnants have small transverse momenta, as shown in the
sketch. In rare events, there is a hard parton-parton collision, which
produces jets with high transverse momenta. In the event shown, there
are two high
$P_T$ jets.}
\label{twojets}
\end{figure}

These ideas work for hadron-hadron collisions too. In such
collisions, there is sometimes a hard parton-parton collision, which
produces two or more jets, as depicted in Fig.~\ref{twojets}. Consider
the cross section to make one jet plus anything else, 
\begin{equation}
A + B \to jet + X.
\end{equation}
Let $E_T$ be the {\it transverse energy} of the jet, defined as the
sum of the absolute values of the transverse momenta of the
particles in the jet. Let $y$ be the rapidity of the jet. Given a
definition of exactly what it means to have a jet with transverse
energy $E_T$ and rapidity $y$, the jet production cross section
takes the familiar factored form
\begin{eqnarray}
{\green{d \sigma \over d E_T d\eta}} &\approx& 
\sum_{a,b} \int_{x_A}^1\! d \xi_A \int_{x_B}^1\! d \xi_B\
{\blue{f_{a/A}(\xi_A,\mu_F)}}\ {\blue{f_{b/B}(\xi_B,\mu_F)}}\
{\red{{{d \hat\sigma^{ab}(\mu,\mu_F) \over d E_T d\eta}}}}.
\end{eqnarray}

\begin{figure}[htb]
\centerline{\DESepsf(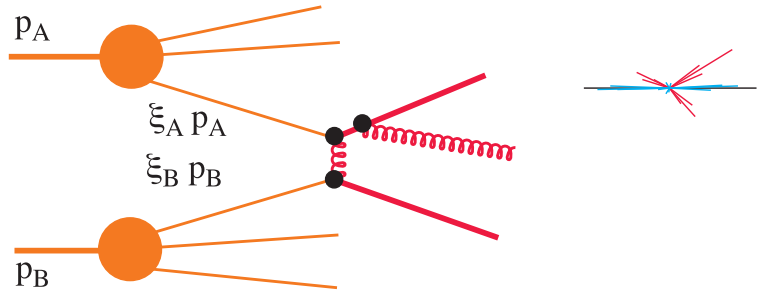 width 10 cm)}
\caption{A Feynman diagram for jet production in hadron-hadron
collisions. The leading order diagrams for $A + B \to jet + X$ occur
at order $\alpha_s^2$. This particular diagram is for an interaction
of order $\alpha_s^3$. When the emitted gluon is not soft or nearly
collinear to one of the outgoing quarks, this diagram corresponds to
a final state like that shown in the small sketch, with three jets
emerging in addition to the beam remnants. Any of these jets can be
the jet that is measured in the one jet inclusive cross section.}
\label{jet}
\end{figure}

What shall we choose for the definition of a jet? At a crude level,
high $E_T$ jets are quite obvious and the precise definition hardly
matters. However, if we want to make a quantitative measurement of a
jet cross section to compare to next-to-leading order theory, then the
definition does matter. There are several possibilities for a
definition that is infrared safe. The one most used in
hadron-hadron collisions is based on cones.

\begin{figure}[htb]
\centerline{\DESepsf(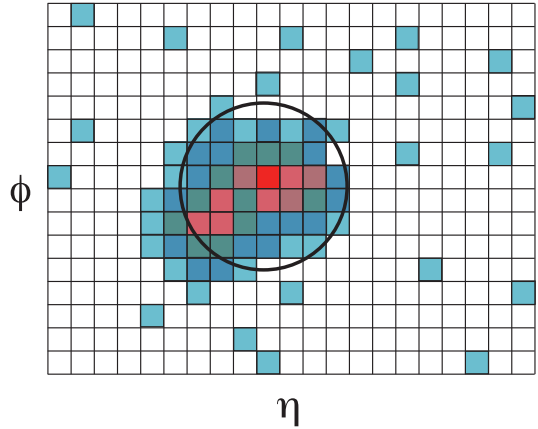 width 7 cm)}
\caption{Jet definition according to the Snowmass algorithm. The
shading of the squares represents the density of transverse energy as
a function of azimuthal angle $\phi$ and pseudorapidity $\eta$. The
cells inside the circle constitute the jet.}
\label{jetcone}
\end{figure}

In the standard {\it Snowmass Accord} definition \cite{snowmass}, one
imagines that the experimental calorimeter is divided into small
angular cells labeled $i$ in $\eta$-$\phi$ space, as depicted in
Fig.~\ref{jetcone}. We can say that a jet consists of all the
particles that fall into certain of the calorimeter cells, or we can
measure the $E_T$ in each cell and build the jet parameters from the
cell variables $(E_{Ti},\eta_i,\phi_i)$. We then say that a jet
consists of the cells inside a certain circle in $\eta$-$\phi$ space.
The circle has a radius $R$, usually chosen as 0.7 radians, and is
centered on a direction $(\eta_J,\phi_J)$. Thus the calorimeter cells
$i$ included in the jet obey
\begin{equation}
(\eta_i - \eta_J)^2 + (\phi_i - \phi_J)^2 < R^2.
\end{equation}
The transverse energy of the jet is defined to be
\begin{equation}
E_{T,J} = \sum_{i\in {\rm cone}}E_{T,i}.
\label{jetET}
\end{equation}
The direction of the jet is defined to be the direction
$(\eta_J,\phi_J)$ of the jet axis, which is chosen to obey 
\begin{equation}
\phi_J = {1 \over E_{T,J}}\sum_{i\in {\rm cone}}E_{T,i}\ \phi_i,
\label{jetphi}
\end{equation}
\begin{equation}
\eta_J = {1 \over E_{T,J}}\sum_{i\in {\rm cone}}E_{T,i}\ \eta_i
\label{jeteta}
\end{equation}
Of course, if one picks a trial jet direction $(\eta_J,\phi_J)$ to
define the meaning of ``${i\in {\rm cone}}$'' and then computes
$(\eta_J,\phi_J)$ from these equations, the output jet direction will
not necessarily match the input cone axis. Thus one has to treat the
equations iteratively until a consistent solution is found.

Note that the Snowmass algorithm for computing
$E_{T,J},\phi_J,\eta_J$ is infrared safe. Infinitely soft particles
do not affect the jet parameters because they enter the equations
with zero weight. If two particles have the same angles
$\eta,\phi$, then it does not matter if we join them together into
one particle before applying the algorithm. For example
\begin{equation}
E_{T,1} \phi + E_{T,2} \phi = 
(E_{T,1} + E_{T,2})\, \phi.
\end{equation}

Note, however, that the Snowmass definition given above is not
complete. It is perfectly possible for two or more cones that are
solutions to  Eqs.~(\ref{jetET},\ref{jetphi},\ref{jeteta}) to
overlap. One must then have an algorithm to assign calorimeter cells
to one of the competing jets, thus splitting the jets, or else to
merge the jets. When supplemented by an appropriate split/merge
algorithm, the Snowmass definition is not as simple as it seemed at
first.

In an order $\alpha_s^3$ perturbative calculation, one simply
applies this algorithm at the parton level. At this order of
perturbation theory, there are two or three partons in the final
state. In the case of three partons in the final state, two of them
are joined into a jet if they are within $R$ of the jet axis
computed from the partonic momenta. The split/merge question does
not apply at this order of perturbation theory.

I showed a comparison of the theory and experiment for the one jet
inclusive cross section in Fig.~\ref{Jetcteq3}.

I should record here that the actual jet definitions used in current
experiment are close to the Snowmass definition given above but are
not exactly the same. Furthermore, there are other definitions
available that may come into use in the future. There is not time
here to explore the issues of jet definitions in detail. What I hope
to have done is to give the outline of one definition and to
explore what the issues are.

\section{Epilogue}

QCD is a rich subject. The theory and the experimental evidence
indicate that quarks and gluons interact weakly on short time and
distance scales. But the net effect of these interactions extending
over long time and distance scales is that the chromodynamic force is
strong. Quarks are bound into hadrons. Outgoing partons emerge as
jets of hadrons, with each jet composed of subjets. Thus QCD theory
can be viewed as starting with simple perturbation theory, but it does
not end there. The challenge for both theorists and experimentalists
is to extend the range of phenomena that we can relate to the
fundamental theory.

\medskip
I thank F.\ Hautmann for reading the manuscript and helping to
eliminate some of the mistakes. 


\end{document}